\documentclass[12pt]{article}
\usepackage{jheppub}

\pdfoutput=1

\usepackage{amsmath,bbm,array,amsfonts,graphicx,wrapfig,lscape,float,mathtools,multirow,longtable}
\usepackage[dvipsnames,table]{xcolor}
\usepackage{array}
\usepackage{color}

\newcommand{\be}{\begin{equation}}
\newcommand{\ee}{\end{equation}}
\newcommand{\beq}{\begin{equation}}
\newcommand{\beql}[1]{\begin{equation}\label{#1}}
\newcommand{\eeq}{\end{equation}}
\newcommand{\ba}{\begin{array}}
\newcommand{\ea}{\end{array}}
\newcommand{\bea}{\begin{eqnarray}}
\newcommand{\beal}[1]{\begin{eqnarray}\label{#1}}
\newcommand{\eea}{\end{eqnarray}}
\newcommand{\ben}{\begin{enumerate}}
\newcommand{\een}{\end{enumerate}}
\newcommand{\bean}{\begin{eqnarray*}}
\newcommand{\eean}{\end{eqnarray*}}
\newcommand{\eref}[1]{(\ref{#1})}
\newcommand{\sref}[1]{\S\ref{#1}}
\newcommand{\tref}[1]{Table~\ref{#1}}
\newcommand{\nn}{\nonumber}

\newcommand{\fref}[1]{Figure \ref{#1}}
\newcommand{\btab}[1]{\begin{tabular}{#1}}
\newcommand{\etab}{\end{tabular}}

\def \Black {\pdfsetcolor{0 0 0 1}}

\newcommand{\comment}[1]{}

\newcommand{\IC}{\mathbb{C}}

\newcommand{\ud}{\mathrm{d}}

\newcommand{\qed}{\nobreak \ifvmode \relax \else
      \ifdim\lastskip<1.5em \hskip-\lastskip
      \hskip1.5em plus0em minus0.5em \fi \nobreak
      \vrule height0.75em width0.5em depth0.25em\fi}

\definecolor{darkspringgreen}{rgb}{0.09, 0.45, 0.27}
\definecolor{forestgreen}{rgb}{0.13, 0.55, 0.13}

\usepackage{array}
\usepackage{physics}
\usepackage{subcaption}
\usepackage{tikz,tikz-3dplot}
\usepackage[colorlinks=true]{hyperref}
\newcolumntype{C}[1]{>{\centering\let\newline\\\arraybackslash\hspace{0pt}}m{#1}}

\definecolor{yellow2}{rgb}{0.98, 0.80, 0.20}

\definecolor{forestgreen}{rgb}{0.13, 0.55, 0.13}
\definecolor{ceruleanblue}{rgb}{0.16, 0.32, 0.75}

\definecolor{darkspringgreen}{rgb}{0.09, 0.45, 0.27}
\definecolor{forestgreen}{rgb}{0.13, 0.55, 0.13}

\newcommand{\cC}{{\cal C}}

\title{Mass Deformations of Brane Brick Models}

\author[a,b,c]{Sebasti\'an Franco,}
\author[d]{Dongwook Ghim,}
\author[a,b]{Georgios P. Goulas,}
\author[e,f]{\\ and Rak-Kyeong Seong}

\affiliation[a]{Physics Department, The City College of the CUNY\\
	160 Convent Avenue, New York, NY 10031, USA}
	
\affiliation[b]{Physics Program and \textsuperscript{$c$}Initiative for the Theoretical Sciences\\
	The Graduate School and University Center, The City University of New York\\
	365 Fifth Avenue, New York NY 10016, USA}

\affiliation[d]{
Yukawa Institute for Theoretical Physics, Kyoto University, \\
Sakyo-ku, Kyoto 606-8502, Japan
}
	
\affiliation[e]{
Department of Mathematical Sciences, and
\textsuperscript{$f$}Department of Physics,\\
Ulsan National Institute of Science and Technology,\\
50 UNIST-gil, Ulsan 44919, South Korea
}

\emailAdd{sfranco@ccny.cuny.edu}
\emailAdd{dghim@yukawa.kyoto-u.ac.jp}
\emailAdd{ggoulas@gradcenter.cuny.edu}
\emailAdd{seong@unist.ac.kr}

\preprint{
\begin{flushright}
UNIST-MTH-23-RS-02 \\
YITP-23-47\\
\end{flushright}
}

\abstract{
We investigate a class of mass deformations that connect pairs of $2d$ $(0,2)$ gauge theories associated to different toric Calabi-Yau 4-folds. These deformations are generalizations to $2d$ of the well-known Klebanov-Witten deformation relating the $4d$ gauge theories for the $\mathbb{C}^2/\mathbb{Z}_2 \times \mathbb{C}$ orbifold and the conifold. We investigate various aspects of these deformations, including their connection to brane brick models and the relation between the change in the geometry and the pattern of symmetry breaking triggered by the deformation. We also explore how the volume of the Sasaki-Einstein 7-manifold at the base of the Calabi-Yau 4-fold varies under deformation, which leads us to conjecture that it quantifies the number of degrees of freedom of the gauge theory and its dependence on the RG scale.}
\begin{document}

\maketitle

\section{Introduction}

An infinite class of $2d$ $(0,2)$ supersymmetric gauge theories can be realized on the worldvolume of D1-brane probing toric Calabi-Yau 4-folds. 
The field content and interactions of these gauge theories is beautifully described by a tessellation of a 3-torus, known as the brane brick model \cite{Franco:2015tna,Franco:2015tya}.
3-dimensional brane bricks, 2-dimensional brick faces and 1-dimensional brick edges in the brane brick model represent $U(N)$ gauge groups, bifundamental or adjoint chiral fields and Fermi fields, and $J$- and $E$-terms, respectively. 
Brane brick models simplify the correspondence between the $2d$ supersymmetric gauge theory and the corresponding toric Calabi-Yau geometry. 
Since the introduction of brane brick models in \cite{Franco:2015tna,Franco:2015tya}, they have been studied for example in the context of Calabi-Yau mirror symmetry \cite{Franco:2016qxh,Franco:2016tcm}, their relationship to $4d$ $\mathcal{N}=1$ supersymmetric gauge theories \cite{Franco:2016fxm,Franco:2018qsc, Franco:2021ixh} represented by brane tilings \cite{Hanany:2005ve, Franco:2005rj}, and in relationship to various classes of toric Calabi-Yau 4-folds \cite{Franco:2022gvl,Franco:2022isw}.

In this work, we investigate how relevant deformations such as mass deformations on the $2d$ $(0,2)$ supersymmetric gauge theory affect the corresponding brane brick model and toric Calabi-Yau 4-fold. 
The study of such mass deformations was initiated many years ago in the context of $4d$ supersymmetric gauge theories by the seminal work by Klebanov and Witten in \cite{Klebanov:1998hh}, where a certain mass deformation of a $4d$ supersymmetric gauge theory corresponding to an orbifold of the form $\mathbb{C}^2/\mathbb{Z}_2 \times \mathbb{C}$ led to the construction of the conifold theory \cite{Candelas:1989ug, Candelas:1989js}. 
With the systematic construction and classification of $4d$ $\mathcal{N}=1$ supersymmetric gauge theories corresponding to toric Calabi-Yau 3-folds through the introduction of brane tilings \cite{Hanany:2005ve, Franco:2005rj, Hanany:2012hi}, the $4d$ analogues to brane brick models, it became a natural problem to investigate mass deformations of brane tilings representing large classes of $4d$ $\mathcal{N}=1$ supersymmetric gauge theories \cite{Bianchi:2014qma}.

We initiate a systematic investigation of mass deformations on $2d$ $(0,2)$ gauge theories with a D-brane realization, focusing on the class of deformations for which both the initial and final theories are associated to toric Calabi-Yau 4-folds. The first example of such a deformation was presented in \cite{Franco:2016fxm}. We do this by specifying how adding certain mass terms to the $J$- and $E$-terms ensures that after mass deformation the resulting $2d$ $(0,2)$ supersymmetric gauge theory can be again represented by a brane brick model with an associated toric Calabi-Yau 4-fold. 
We further investigate how mass deformations affect the combinatorial structure of brane brick models given by brick matchings \cite{Franco:2015tya}.
Such brick matchings correspond to GLSM fields \cite{Witten:1993yc} that are used to describe the toric Calabi-Yau 4-fold associated to the brane brick model. 
Moreover, exploiting how mass deformations affect brick matchings, we are able to determine how certain global symmetries of brane brick models are broken and how the toric Calabi-Yau 4-fold is affected. This geometric deformation is measured by the change of minimum volume of the corresponding Sasaki-Einstein 7-manifold \cite{Martelli:2005tp, Martelli:2006yb, He:2017gam}, which forms the base manifold of the toric Calabi-Yau 4-fold. 
Our work presents several examples for mass deformations of brane brick models and paves the way towards a full classification of possible deformations of brane brick models. 

Our work is organized as follows.
In section \sref{sec:2}, we give a brief review on brane brick models and how brick matchings corresponding to GLSM fields are used to describe the corresponding toric Calabi-Yau 4-folds. 
We further describe the different moduli spaces that are associated to brane brick models and how the connect to the different global symmetries of the corresponding $2d$ $(0,2)$ supersymmetric gauge theories. 
In section \sref{sec:3}, we introduce mass deformations for brane brick models and their effect on brick matchings, the geometry of the toric Calabi-Yau 4-folds and the global symmetries of the brane brick model. 
By calculating the minimum volumes the Sasaki-Einstein base manifolds of the toric Calabi-Yau 4-folds using the Hilbert series, we also determine how the minimum volume changes under mass deformation of brane brick models. 
We give several explicit examples of mass deformation for brane brick models in section \sref{sec:4}. We present our conclusions and future directions in section \sref{sec:5}.
\\

\section{Background \label{sec:2}}

Since the introduction of brane brick models, multiple methods of connecting a given toric Calabi-Yau 4-fold with a corresponding $2d$ $(0,2)$ gauge theory have been introduced \cite{Franco:2015tna, Franco:2015tya, Franco:2016tcm, Franco:2016qxh, Franco:2016fxm,Franco:2016nwv,Franco:2017cjj, Franco:2018qsc}. In all these methods, it can be seen that brane brick models tremendously simplify the correspondence between the Calabi-Yau geometry and the $2d$ $(0,2)$ gauge theory. 
One of the key ingredients in this correspondence is a combinatorial object known as a \textit{brick matching} \cite{Franco:2015tya}. 

\subsection{Brick Matchings and Geometry \label{sec:21}}

In the following section, we will give a brief summary of how brick matchings appear in brane brick models, and what role they play in the Calabi-Yau 4-fold geometry as well as in the corresponding $2d$ $(0,2)$ gauge theory.

\paragraph{Brick Matchings.}
Brick matchings are combinatorial objects defined through the $J$- and $E$-terms of the brane brick model, which take the following binomial form 
\beal{es05a01}
\Lambda_{ij} &~:~&
J_{ji} = J_{ji}^{+} - J_{ji}^{-} 
\nn\\
\bar{\Lambda}_{ij} &~:~&
E_{ij} = E_{ij}^{+} - E_{ij}^{-}  
~,~
\eea
which is often referred to as the toric condition. In order to define the brick matchings, we complete the $J$- and $E$-terms by multiplying them by the corresponding Fermi fields $\Lambda_{ij}$ or $\bar{\Lambda}_{ij}$.
This results in two pairs of monomial terms known as \textit{plaquettes} for every $(\Lambda_{ij}, \bar{\Lambda}_{ij})$-pair, 
\beal{es05a02}
\Lambda_{ij} \cdot J_{ji}^{+} ~,~
\Lambda_{ij} \cdot J_{ji}^{-} ~,~
\bar{\Lambda}_{ij} \cdot E_{ij}^{+} ~,~
\bar{\Lambda}_{ij} \cdot E_{ij}^{-} ~,~
\eea
where $J_{ji}^{\pm}$ and $E_{ij}^{\pm}$ indicate monomial products of chiral fields. 
Given plaquettes, \textit{brick matchings} can be defined as a special collection of chiral, Fermi and conjugate Fermi fields that cover every plaquette exactly once by satisfying the following conditions:
\begin{itemize}
\item The chiral fields in the brick matching cover the plaquettes $(\Lambda_{ij} \cdot J_{ji}^{+},~ \Lambda_{ij} \cdot J_{ji}^{-} )$ or the plaquettes $(\bar{\Lambda}_{ij} \cdot E_{ij}^{+},~ \bar{\Lambda}_{ij} \cdot E_{ij}^{-} )$ exactly once each. 

\item If the chiral fields in the brick matching cover the plaquettes  $(\Lambda_{ij} \cdot J_{ji}^{+},~ \Lambda_{ij} \cdot J_{ji}^{-} )$, then $\bar{\Lambda}_{ij}$ is included in the brick matching.

\item If the chiral fields in the brick matching cover the plaquettes $(\bar{\Lambda}_{ij} \cdot E_{ij}^{+},~ \bar{\Lambda}_{ij} \cdot E_{ij}^{-} )$, then $\Lambda_{ij}$ is included in the brick matching. 
\end{itemize}
The chiral fields $X_m$ contained in brick matching $p_\mu$ can be summarized in a brick matching matrix $P$, whose entries take the following form,
\beal{es05a10}
P_{m\mu} =  \left\{ \ba{cc}
1 & ~~~X_m \in p_\mu \\
0 & ~~~X_m \notin p_\mu 
\ea
\right.
~.~
\eea
Given the $P$-matrix, the chiral fields of a brane brick model can be expressed in terms of brick matchings as follows
\beal{es05a11}
X_m = \prod_\mu p_\mu^{P_{m\mu}}
~.~
\eea

We can extend the definition of the $P$-matrix in \eref{es05a10}, such that it includes the Fermi field content of each brick matching as follows,
\beal{es05a26}
&
P_{\Lambda,X_m p_\mu}
=  \left\{ \ba{cc}
1 & ~~~X_m \in p_\mu \\
0 & ~~~X_m \notin p_\mu \\
\ea
\right.
~~,~~
P_{\Lambda,\Lambda_m p_\mu}
=  \left\{ \ba{cc}
1 & ~~~\Lambda_m \in p_\mu \\
1 & ~~~\bar{\Lambda}_m \in p_\mu \\
0 & ~~~\text{otherwise} \\
\ea
\right.
&
~.~
\eea

Brick matchings can be interpreted as GLSM fields in the toric description of the mesonic moduli space.

\paragraph{Forward Algorithm and GLSM Fields.}
The mesonic moduli space is the toric Calabi-Yau 4-fold associated to the brane brick model and can be obtained through the \textit{forward algorithm} \cite{Franco:2015tya}. 
In terms of brick matchings $p_1, \dots, p_c$ representing GLSM fields, the mesonic moduli space can be expressed as the following symplectic quotient \cite{Franco:2005sm, Forcella:2008eh, Franco:2015tna},
\beal{es05a15}
\mathcal{M}^{mes} =\left( \mathbb{C}^c[p_1, \dots, p_c] // Q_{JE} \right) // Q_{D}
~,~
\eea
where the $Q_{JE}$ are the $U(1)$ charges originating from the $J$- and $E$-terms, and $Q_D$ are the $U(1)$ charges originating from the D-terms of the brane brick model.

The $Q_{JE}$ charge matrix can be obtained directly from the brick matching matrix in \eref{es05a10} following
\beal{es05a16}
Q_{JE} = \ker P ~,~
\eea
whereas the $Q_D$-matrix is obtained following
\beal{es05a17}
\Delta = Q_D \cdot P^{T} ~,~
~,~
\eea
where $\Delta$ is the reduced incidence matrix of the quiver corresponding to the brane brick model and $P^T$ is the transpose of the $P$-matrix in \eref{es05a10}. 

The toric diagram in terms of GLSM fields is given by the following coordinate matrix
\beal{es05a18}
G_t = \ker 
\left(
\ba{c}
Q_{JE} \\
Q_D
\ea
\right) ~,~
\eea
where columns of the $G_t$-matrix correspond to the coordinates of points in the 3-dimensional toric diagram of the Calabi-Yau 4-fold. 
These points in the toric diagram correspond to brick matchings in the associated brane brick model. 

\begin{figure}[H]
\begin{center}
\resizebox{0.35\hsize}{!}{
\includegraphics[height=6cm]{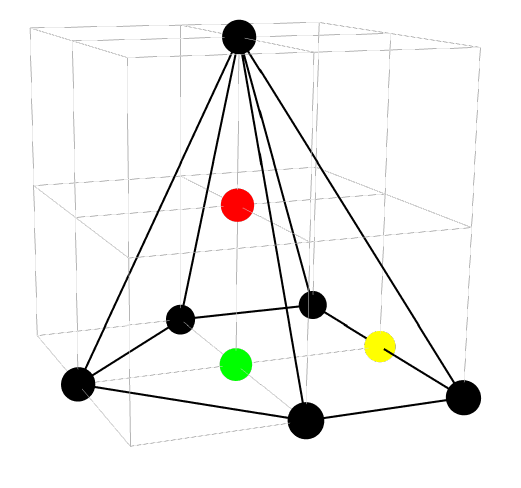} 
}
\caption{
The points in the 3-dimensional toric diagram of a toric Calabi-Yau 4-fold can be classified into extremal points (black), points internal on faces (green), points internal on edges (yellow) and points in the interior of the toric diagram (red). 
\label{f_extremalpoints}}
 \end{center}
 \end{figure}

Toric diagrams of toric Calabi-Yau 4-folds are convex lattice polytopes in $3$ dimensions and points in the toric diagram can be classified into \textit{extremal points}, which correspond to corner points of the toric diagram.
All other points are not extremal and may be internal on edges or faces of the toric diagram, or points in the 3-dimensional interior of the toric diagram as illustrated in \fref{f_extremalpoints}.
Extremal points in the toric diagram are associated always to a single GLSM fields corresponding to a single brick matching, whereas internal points have a multiplicity and are associated with multiple GLSM fields corresponding to multiple brick matchings. 
We call the brick matchings corresponding to extremal points in the toric diagram as \textit{extremal brick matchings}.

It was observed in \cite{Franco:2015tna} that the forward algorithm for brane brick models sometimes leads to \textit{extra GLSM fields}.
These extra GLSM fields appear as additional points in the toric diagram that are positioned outside the 3-dimensional Calabi-Yau hyperplane.
It was shown in \cite{Franco:2015tna} that such extra GLSM fields over-parameterize the mesonic moduli space $\mathcal{M}^{mes}$, meaning that the generators and relations of $\mathcal{M}^{mes}$ remain unaffected by the presence of these extra GLSM fields. Extra GLSM fields do not satisfy the brick matching condition. Therefore, whenever we are not considering extra GLSM fields, we will use the terms GLSM field and brick matching interchangeably.

\subsection{Global Symmetries and Brick Matchings \label{sec:22}}

In this section, we give a brief review on global symmetries carried by brick matchings in brane brick models. 
These symmetries relate to the isometries of the mesonic moduli space $\mathcal{M}^{mes}$ \cite{Benvenuti:2004dy,Benvenuti:2006qr, Franco:2015tya,Franco:2015tna} and the master space ${}^{\text{Irr}}\mathcal{F}^\flat$ \cite{Forcella:2008bb,Forcella:2008eh,Forcella:2008ng,Franco:2015tna,Franco:2015tya} of the brane brick model. 

For a gauge theory with $G$ gauge groups, the brick matchings carry $G+3$ charges that relate to the 4 mesonic, including the $U(1)_R$ symmetry, and the $G-1$ baryonic symmetries.\footnote{We simply use the term baryonic as a generalization of the term in the context of CY 3-folds}
The $G-1$ baryonic symmetries originate from the $G$ $U(N)$ gauge groups of the brane brick model, which each give a baryonic $U(1)$ symmetry. Since all chiral and Fermi fields of the brane brick model are bifundamental or adjoint, one of the $U(1)$'s decouples giving a total of $G-1$ baryonic symmetries. 

Accordingly, each brick matching $p_\mu$ is assigned a $(G+3)$-dimensional global symmetry charge vector $\alpha_\mu$ whose components satisfy the following conditions:
\begin{itemize}
\item All $(G+3)$-dimensional charge vectors are linearly independent to each other.
\item The sum of all charge vectors is of the form,
\beal{es06a11}
\sum_{\mu=1}^{c} \alpha_\mu = (0,\dots,0)
~,~
\eea 
where $c$ is the number of brick matchings. For simplicity of the presentation, we will focus on the non-$R$ global symmetries. Extending the argument to include $U(1)_R$ is straightforward.
\end{itemize}

Brick matchings are defined such that when chiral and Fermi fields are expressed in terms of them, every plaquette in the $J$- and $E$-terms contains all the brick matchings. This, combined with the constraint \eqref{es06a11} on the charges of brick matchings, implies that the $J$- and $E$-terms are invariant under the non-$R$ global symmetries and have $R$-charge equal to 1. Conversely, we can interpret \eqref{es06a11} as following from the global symmetry and $R$-charge of the $J$- and $E$-terms. The sum in \eqref{es06a11} can be restricted to extremal brick matchings, since they are the only ones charged under the global symmetries.

Using \eref{es05a26}, the non-$R$ global symmetry charge carried by any chiral or Fermi field can be expressed as follows 
\beal{es07a14}
\alpha(X_m) &=& \sum_{\mu} P_{\Lambda,X_m p_\mu} \alpha_\mu ~,~
\nn\\
\alpha(\Lambda_m) &=& \sum_{\mu} P_{\Lambda, \Lambda_m p_\mu} \alpha_\mu ~,~
\nn\\
\alpha(\bar{\Lambda}_m) &=& \sum_{\mu} (1- P_{\Lambda,\Lambda_m p_\mu}) \alpha_\mu
= - \sum_{\mu} P_{\Lambda,\Lambda_m p_\mu} \alpha_\mu
~,~
\eea
where in the last equation we used the fact that the non-$R$ global symmetry charges sum up to $(0,\dots,0)$ over extremal brick matchings. Of course it agrees with the expectation from conjugation of the Fermis.
\\

\section{Mass Deformations \label{sec:3}}

In this section, we introduce mass deformations of brane brick models and investigate their effect on the corresponding toric geometries. A natural way to understand these deformations is in terms of global symmetries and their breaking. The global symmetries of a $2d$ $(0,2)$ field theory and the charges of fields under them, uniquely determine the corresponding $J$- and $E$-terms. In a standard effective field theory approach, the $J$- and $E$-terms correspond to all the gauge invariant plaquettes consistent with the global symmetry. Any term that is not originally present, including mass terms, would correspond to a breaking of the global symmetry. The specific pattern of symmetry breaking depends on the charges of the deformation term(s). In the class of theories under study, which are engineered using branes at singularities, the underlying toric CY$_4$ determines the global symmetries. Moreover, the charges of perfect matchings are encoded in the linear relationships between the corresponding points in the toric diagram. Therefore, we expect that the addition of deformation terms will modify the global symmetry of the theory, which in term translates into a change in the toric diagram. 
\\

\subsection{Mass Deformations of Brane Brick Models \label{sec:31}}

In this paper, we are interested in mass deformations of brane brick models. In $2d$ $(0,2)$ gauge theories, mass terms are gauge invariant quadratic terms in the Lagrangian involving a chiral-Fermi pair. Therefore, for a theory to admit mass deformations, it must contain pairs of chiral and Fermi fields extending between the same pair of nodes. The mass deformations would then map to linear terms in either $J$- or $E$-terms for the Fermi in the pair.

\paragraph{Mass Deformation.}
Let us introduce the general form of mass deformation for a brane brick model with a quiver diagram $Q$.
The possible mass terms that one can add to the $J$-terms or the $E$-terms take the following general form, 
\beal{es08a01}
(\Lambda_{ij}, X_{ij}) \in Q &~:~& 
J_{ji}^\prime = J_{ji}
~,~
E_{ij}^\prime = \pm m X_{ij} + E_{ij} 
~,~
\nn\\
(\bar{\Lambda}_{ij}, X_{ji}) \in Q &~:~&
J_{ji}^\prime = \pm m X_{ji} + J_{ji}
~,~
E_{ij}^\prime =  E_{ij} 
~,~
\eea
where $J_{ji}$ and $E_{ij}$ are the initial $J$- and $E$-terms of the brane brick model, and
$J_{ji}^\prime$ and $E_{ij}^\prime$ are the $J$- and $E$-terms after mass deformation.

By integrating out the Fermi fields $\Lambda_{ij}$, the above modifications of the $J$- and $E$-terms in \eref{es08a01} respectively lead to the following chiral field replacements,
\beal{es08a02}
X_{ij} = \mp \frac{1}{m} (E^+_{ij} - E^-_{ij}) ~~~~,~~~~
X_{ji} = \mp \frac{1}{m} (J^+_{ji} - J^-_{ji}) ~,~
\eea
in the remaining $J$- and $E$-terms of the brane brick model.

\paragraph{Toric Condition.}
We note that before and after mass deformation, the $J$- and $E$-terms have to satisfy the \textit{toric condition} \cite{Feng:2002zw, Franco:2015tna,Franco:2015tya} of brane brick models.
This means that both before and after mass deformation, the $J$- and $E$-terms have to take the form of binomial relations,
\beal{es08a03}
\Lambda_{ij} ~:~
J_{ji} = J_{ji}^{+} - J_{ji}^{-} 
~~~~,~~~~~
\bar{\Lambda}_{ij} ~:~
E_{ij} = E_{ij}^{+} - E_{ij}^{-}  
~,~
\eea
where $J^\pm_{ji}$ and $E^\pm_{ij}$ are holomorphic monomials in chiral fields. 
In order to satisfy the toric condition even after mass deformation, there are various ways to adjust the mass deformation in \eref{es08a01} as follows:
\begin{itemize}
\item \underline{Sign of the Mass Term.}
The signs in the mass terms 
\beal{es08a05}
\Delta J_{ji} = \pm m X_{ji} 
~~~~,~~~~
\Delta E_{ij} = \pm m X_{ij} 
~,~
\eea
in \eref{es08a01} can be chosen in such a way that 
additional terms generated by the chiral field replacements in \eref{es08a02}
cancel each other such that the resulting $J$- and $E$-terms remain binomial and satisfy the toric condition.

\item \underline{Pairs of Mass Terms.}
We add mass terms in pairs for either a pair of $J$-terms or a pair of $E$-terms. 
For example, if the quiver of the brane brick model contains a pair of chiral-Fermi pairs of the form $(\Lambda_{ij}, X_{ij})$ and $(\Lambda_{kl}, X_{kl})$, we can add a pair of mass terms to the $E$-terms as follows, 
\beal{es08a06}
(\Lambda_{ij}, X_{ij}) \in Q &~:~& 
J_{ji}^\prime = J_{ji}
~,~
E_{ij}^\prime = + m X_{ij} + E_{ij} 
~,~
\nn\\
(\Lambda_{kl}, X_{kl}) \in Q &~:~& 
J_{lk}^\prime = J_{lk}
~,~
E_{kl}^\prime = - m X_{kl} + E_{kl} 
~,~
\eea
where the mass terms are chosen to have opposite signs.
Similarly, if there is a pair of chiral-Fermi pairs of the form $(\bar{\Lambda}_{ij}, X_{ji})$ and $(\bar{\Lambda}_{kl}, X_{lk})$, we can add the following pair of mass terms to the $J$-terms as follows,
\beal{es08a07}
(\bar{\Lambda}_{ij}, X_{ji}) \in Q &~:~&
J_{ji}^\prime = + m X_{ji} + J_{ji}
~,~
E_{ij}^\prime =  E_{ij} 
~,~
\nn\\
(\bar{\Lambda}_{kl}, X_{lk}) \in Q &~:~&
J_{lk}^\prime = - m X_{lk} + J_{lk}
~,~
E_{kl}^\prime =  E_{kl} 
~,~
\eea
where the mass terms are again chosen to have opposite signs.
Adding pairs of mass terms with opposite sign to pairs of $J$-terms and $E$-terms
ensures the cancellation of additional terms that are generated by the chiral field replacements in \eref{es08a02}. The need for mass deformation terms with opposite signs is indeed common in the $4d$ $\mathcal{N}=1$ gauge theories associated to toric CY 3-folds \cite{Klebanov:1998hh,Bianchi:2014qma}.

\item \underline{Redefinition of Interaction Terms.}
We note that the Lagrangian of the $2d$ $(0,2)$ supersymmetric gauge theory has the following $(0,2)$ interaction term,
\beal{es08a10}
L_J = - \int d^2 y ~d^2\theta^+ \sum_a \left(
\Lambda_a J_a (\Phi_i)|_{\bar{\theta}^+=0}
\right)
- h.c. ~,~
\eea
where $J_a (\Phi_i)$ is a $J$-term in terms of chiral fields $\Phi_i$.\footnote{To simplify this general discussion, we use a different notation than in previous sections, with a single subindices $a$ and $i$ to label Fermis and chirals, respectively.} In certain cases, the mass-deformed $J$-terms and $E$-terms can be made to satisfy the toric condition under redefinitions of the above interaction terms.
Such redefinitions can take the following form,
\beal{es08a11}
\Lambda_{ij}^\prime \cdot X_{jk}^\prime 
= 
\Lambda_{ij} \cdot  (X_{jk} + \sum_{h} c_h^{(jk)} X_{jh} X_{hk})
~,~
\eea
with an appropriate choice of coefficients $c_h^{(jk)}$.
We note that redefining a single Fermi-chiral interaction term $\Lambda_{ij}^\prime \cdot X_{jk}^\prime$ is different to redefining a single chiral field with $X_{jk}^\prime =X_{jk} + \sum_{h} c_h^{(jk)} X_{jh} X_{hk}$, as it is done for mass deformations for $4d$ $\mathcal{N}=1$ supersymmetric gauge theories given by brane tilings \cite{Bianchi:2014qma}. 
This is because a redefinition of chiral fields generates multiple non-toric $J$- and $E$-terms rather than eliminating terms that violate the toric condition.
The redefinition of single interaction terms as shown in \eref{es08a11} allows us to target specific non-toric interaction terms. 

\end{itemize}

In section \sref{sec:4}, we present explicit examples illustrating the methods discussed above.

\begin{figure}[H]
\begin{center}
\resizebox{0.95\hsize}{!}{
\includegraphics[height=6cm]{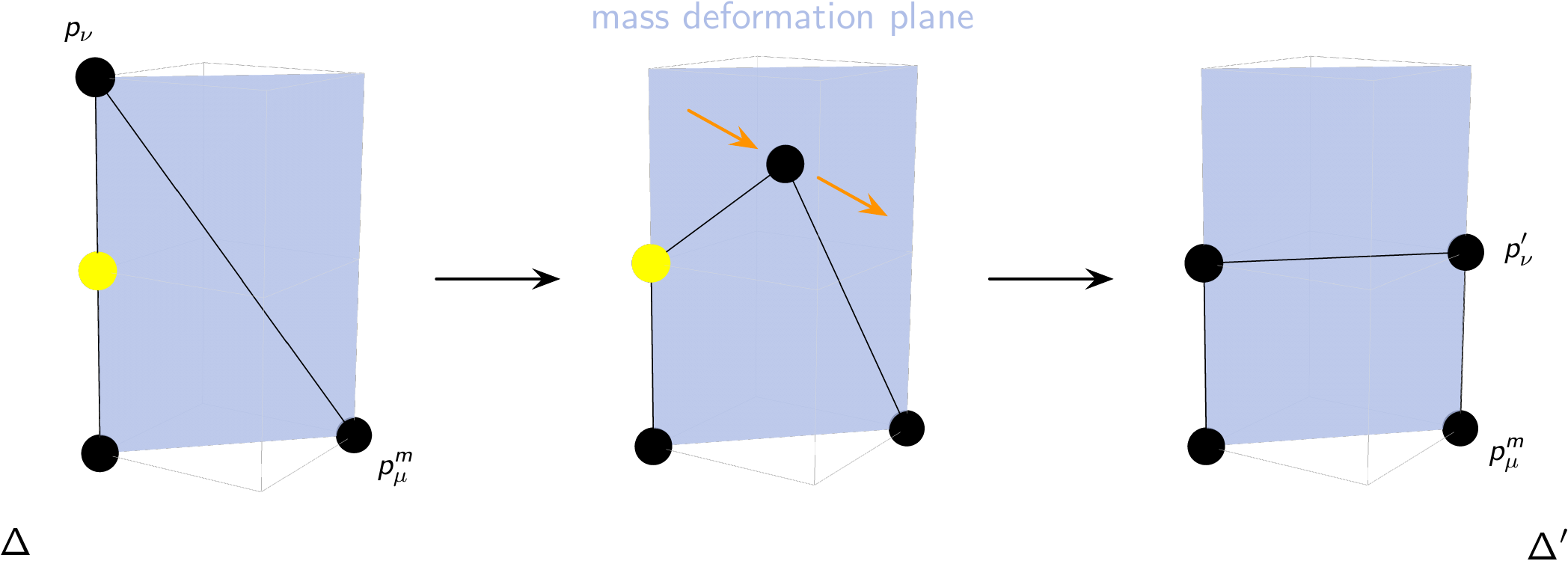} 
}
\caption{
The two end points of the edge $(p_\mu^m,p_\nu)$ correspond to the brick matching $p_\mu^m$ that becomes massive and to the brick matching $p_\nu$ whose toric point changes position under mass deformation becoming brick matching $p'_\nu$. 
We note that the toric points corresponding to $(p_\mu^m,p_\nu, p'_\nu)$ are on a 3-dimensional plane, which we call the \textit{mass deformation plane}.
\label{f_deform0}}
 \end{center}
 \end{figure}

\begin{figure}[H]
\begin{center}
\resizebox{0.95\hsize}{!}{
\includegraphics[height=6cm]{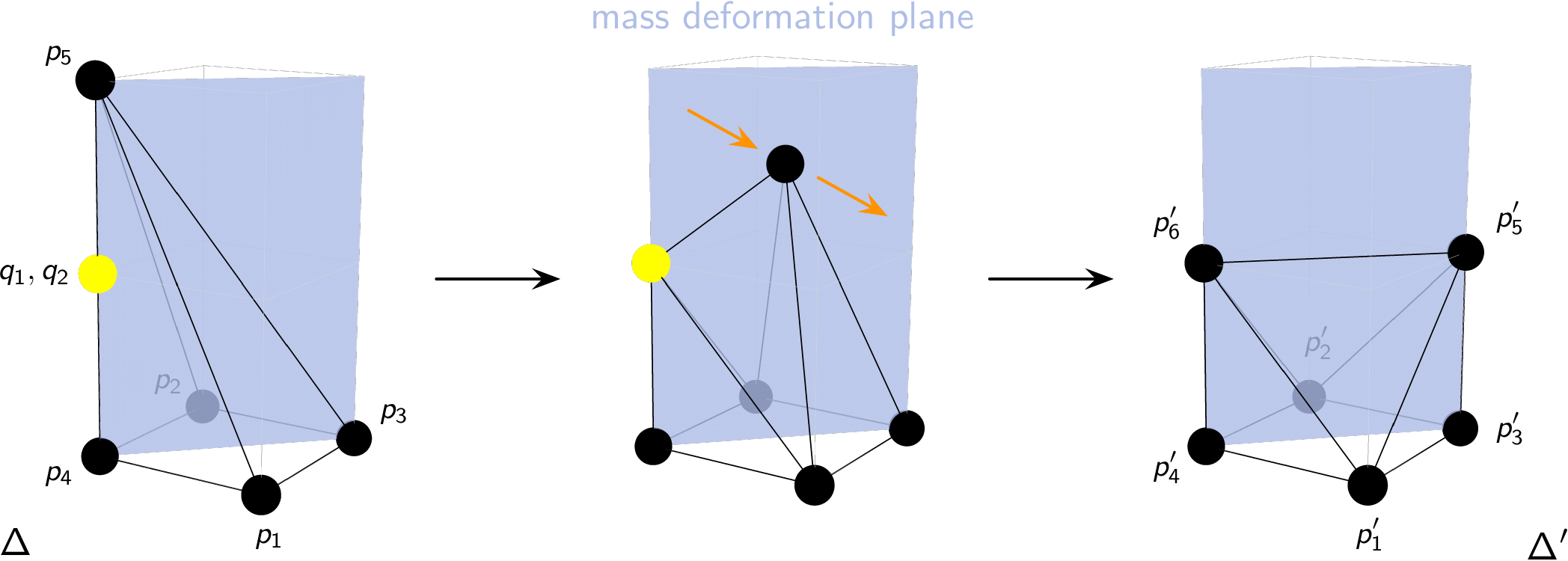} 
}
\caption{
The toric diagram for $\mathcal{C}_{++}$, which becomes the toric diagram for $P^1_{++}(\mathcal{C})$ under the mass deformation of the corresponding brane brick model. The brick matching $p_3$ becomes massive and the toric point corresponding to $p_5$ moves to a new position along the mass deformation plane under mass deformation.
\label{f_deform}}
 \end{center}
 \end{figure}

\paragraph{Effect on the Toric Diagram.}
As discussed in section \sref{sec:22}, brick matchings carry charges corresponding to the global symmetries of the associated $2d$ $(0,2)$ supersymmetric gauge theory. 
Keeping the connection between brick matchings and global symmetries of the brane brick model in mind, we now comment on the effect of mass deformations on brick matchings in the toric diagram.The explicit examples presented in \sref{sec:4} will illustrate all these points. We will see that, as for the $4d$ $\mathcal{N}=1$ theories associated to toric CY 3-folds, pairs of fields that can become massive arise when edges in the toric diagram contain more than two collinear points.\footnote{It would be interesting to have a mathematical proof of this fact and to determine whether this is the only case in which such pairs of fields can arise.} Under mass deformations, a brane brick model exhibits three types of brick matchings, which we summarize below:

\begin{itemize}
\item \underline{Massive Brick Matchings:} 
Brick matchings that contain \textit{all} massive chiral fields in a mass deformation are called \textit{massive brick matchings}.
We recall that for a chiral-Fermi pair of the form $(\Lambda_{ij}, X_{ij})$ with mass term $\Delta E_{ij} = \pm m X_{ij}$, the massive field $X_{ij}$ is part of a corresponding massive brick matching $p_\mu^m$, i.e. $X_{ij} \in p_\mu^m$.
For the case when the mass term takes the form $\Delta J_{ji} = \pm m X_{ji}$ with $(\bar{\Lambda}_{ij}, X_{ji})$, the corresponding massive field $X_{ji}$ is part of a massive brick matching $p_\mu^m$, i.e. $X_{ji} \in p_\mu^m$.
We note that although massive, these brick matchings $p_\mu^m$ correspond to extremal points in the toric diagram whose coordinates do not change relative to the other points in the toric diagram under mass deformation (modulo $SL(3,\mathbb{Z})$ transformations). 
Because these brick matchings contain chiral fields that become massive under mass deformation, the field content of massive brick matchings changes under the deformation. 

\item \underline{Moving Brick Matchings:} 
These extremal brick matchings $p_\mu$ do \textit{not} contain chiral fields that become massive under the mass deformation.
However, under mass deformation, the associated extremal points in the toric diagram change their coordinates relative to the other points in the toric diagram.
Moreover, because these \textit{moving brick matchings} are extremal brick matchings, they can be considered end points of side edges of the toric diagram. 
In fact, a pair consisting of a massive brick matching $p_\mu^m$ and a moving brick matching $p_\nu$ forms prior to the mass deformation the two end points of an external side edge $(p_\mu^m, p_\nu)$ of the original toric diagram $\Delta$, as illustrated in \fref{f_deform0}. By denoting the moving brick matching after mass deformation as $p_\nu^\prime$, we note that the points corresponding to $p_\mu^m$, $p_\nu$ and $p_\nu^\prime$ all lie on a $3$-dimensional plane which intersects both the toric diagram $\Delta$ before the mass deformation and the toric diagram $\Delta^\prime$ after the mass deformation.
We call this plane the \textit{mass deformation plane}.
We illustrate this plane in \fref{f_deform} for the mass deformation of the brane brick model for $\mathcal{C}_{++}$.\footnote{Here $\mathcal{C}$ indicates the conifold. In addition, in this example and the ones that follow, we use the notation introduced in \cite{Franco:2016fxm} for {\it orbifold reduction}, in which the two $+$ subindices mean that a point in the toric diagram of the conifold is lifted twice to produce the toric diagram of a CY$_4$ as shown in \fref{f_deform}.}

We note that the intersections between the plane given by $(p_\mu^m, p_\nu, p_\nu^\prime)$ and $\Delta$ as well as $\Delta^\prime$ give two 2-dimensional convex polygons, which can be considered as toric diagrams of toric Calabi-Yau 3-folds. 
In simple examples discussed in \sref{sec:4}, these 2-dimensional toric diagrams can be associated with brane tilings representing $4d$ $\mathcal{N}=1$ supersymmetric gauge theories that are related by a $4d$ mass deformation \cite{Bianchi:2014qma}.
For example, in \fref{f_deform} we see that the toric diagram on the mass deformation plane corresponds to the abelian orbifold of the form $\mathbb{C}^2/\mathbb{Z}_2 \times \mathbb{C}$ which becomes the toric diagram for the conifold $\mathcal{C}$ after mass deformation. 
The corresponding $4d$ $\mathcal{N}=1$ theories are precisely the two theories originally shown to be related by a Klebanov-Witten flow \cite{Klebanov:1998hh}.

\item \underline{Unaffected Brick Matchings:} 
Besides massive brick matchings containing massive chiral fields and moving brick matchings that correspond to moving extremal points in the toric diagram, there are the remaining brick matchings that are neither massive nor moving. 
These \textit{unaffected brick matchings} can correspond to extremal or non-extremal points in the toric diagram whose coordinates do not change under mass deformation. 
Certain unaffected brick matchings that were originally non-extremal can become extremal brick matchings due to coordinate changes of moving brick matchings. 

\end{itemize}
For a given mass deformation, in the examples in section \sref{sec:4}, we usually observe a single moving brick matching and a single massive brick matching. All other brick matchings stay as unaffected brick matchings during the mass deformation. 

In \fref{f_deform}, we observe that the volume of the toric diagram remains the same, which corresponds to the fact that the number of gauge groups is not affected by mass deformations.

\paragraph{Effect on Global Symmetries.}
Having summarized the effect mass deformation has on brick matchings of a brane brick model, we can now investigate the effect on the global symmetry charges carried by the brick matchings and by extension by the $J$- and $E$-terms of a brane brick model. 

We first note that the plaquettes of the form $\Lambda_{ij} \cdot J^\pm_{ji}$ and $\bar{\Lambda}_{ij} \cdot E^\pm_{ij}$ in a brane brick model are invariant under the global symmetries and therefore carry no global symmetry charge, 
\beal{es08a20}
\alpha(\Lambda_{ij} \cdot J^\pm_{ji}) = (0,\dots,0) ~~,~~
\alpha(\bar{\Lambda}_{ij} \cdot E^\pm_{ij}) = (0,\dots,0) ~,~
\eea
where for the following discussion, we left out the $U(1)$ $R$-charge which is equal to 1 for all $J$- and $E$-term plaquettes. 

Mass deformations correspond to introducing new $J$- or $E$-term plaquettes of the general form 
\beal{es08a21}
m \Lambda_{ij} \cdot X_{ji} ~~,~~
m \bar{\Lambda}_{ij} \cdot X_{ij} ~,~
\eea
which have non-vanishing global symmetry charges,
\beal{es08a22}
\alpha(\Lambda_{ij} \cdot X_{ji}) =
\alpha(\Lambda_{ij}) + \alpha(X_{ji}) 
&=& 
\sum_\mu P_{\Lambda, \Lambda_{ij} p_\mu} \alpha_\mu 
+ \sum_\mu P_{\Lambda, X_{ji} p_\mu} \alpha_\mu 
> 0 
~,~
\nn\\
\alpha(\bar{\Lambda}_{ij} \cdot X_{ij}) =
\alpha(\bar{\Lambda}_{ij} ) + \alpha(X_{ij}) 
&=& 
\sum_\mu (1 - P_{\Lambda, \Lambda_{ij} p_\mu}) \alpha_\mu 
+ \sum_\mu P_{\Lambda, X_{ij} p_\mu} \alpha_\mu 
\nn\\
&=&
- \sum_\mu  P_{\Lambda, \Lambda_{ij} p_\mu} \alpha_\mu 
+ \sum_\mu P_{\Lambda, X_{ij} p_\mu} \alpha_\mu 
> 0 
~,~
\eea
where in the second equation we used the fact that the sum over extremal brick matchings of the non-$U(1)_R$ charges add up to $(0,\dots,0)$, as discussed in section \sref{sec:22}.
The non-zero global symmetry charges on the mass terms break part of the original global symmetry of the brane brick model.

In fact, the charges $\alpha_\mu$ that contribute in \eref{es08a22} correspond to a special set of extremal brick matchings in the brane brick model. 
The associated points of these extremal brick matchings in the 3-dimensional toric diagram of the Calabi-Yau 4-fold are exactly those identified to be on the mass deformation plane.
One of these extremal brick matchings corresponds to a point in the toric diagram that moves along the plane to a new location under mass deformation as illustrated in \fref{f_deform0} and \fref{f_deform}.
The other extremal brick matchings correspond to points that do not change their position along the mass deformation plane. We can therefore interpret the mass deformation plane intersecting the 3-dimensional toric diagram as the directions in the moduli space along which part of the original global symmetry is broken by the mass deformation. 

Given the symmetry breaking pattern triggered by a mass deformation, we can define a new basis of charge vectors of the general form
\beal{es08a23}
\alpha'_\nu = \sum_{\mu} c_{\nu\mu} \alpha_\mu ~,~
\eea
for the new global symmetry after the deformation. This new basis is such that the charges of the mass deformation terms vanish
\beal{es08a25}
\alpha(\Lambda_{ij} \cdot X_{ji}) \Big|_{\alpha_\nu^\prime}  = 0 ~~,~~
\alpha(\bar{\Lambda}_{ij} \cdot X_{ij}) \Big|_{\alpha_\nu^\prime}  = 0  ~.~
\eea
\\

\subsection{Hilbert Series and the Minimum Volume of Sasaki-Einstein 7-Manifolds \label{sec:32}}

In this section, we review Hilbert series of brane brick models and how they can be used to calculate the minimum volume of the Sasaki-Einstein base manifold for the corresponding toric Calabi-Yau 4-fold. We will later investigate how this volume behaves under mass deformations and conjecture its relation to the number of degrees of freedom in the gauge theory. 

\paragraph{Hilbert Series.}
The \textit{Hilbert Series} \cite{Benvenuti:2006qr,Hanany:2006uc,Feng:2007ur} is a generating function for the gauge invariant operators of a supersymmetric gauge theory and has been studied extensively for example in the context of $4d$ $\mathcal{N}=1$ supersymmetric gauge theories given by brane tilings. 
In this work, we make use of the Hilbert series to characterize the mesonic moduli space $\mathcal{M}^{mes}$ of a brane brick model. 
Following the symplectic quotient description of $\mathcal{M}^{mes}$ in section \sref{sec:21}, the corresponding Hilbert series is given by
\beal{es14a01}
g(t_\mu; \mathcal{M}^{mes}) 
= 
\prod_{i=1}^{c-4} \oint_{|z_i|=1} \frac{\ud z_i}{2\pi z_i} \prod_{\mu=1}^{c}
\frac{1}{1-t_\mu \prod_{j=1}^{c-4} z_j^{(Q_t)_{j\mu}}} ~,~
\eea
where $t_\mu$ are the fugacities for the GLSM fields. The number of GLSM fields is given by $c$, and $z_i$ are the fugacities corresponding to the $U(1)$ charges  summarized in $Q_t = (Q_{JE}, Q_D)$.

The Hilbert Series not just characterizes the mesonic moduli space $\mathcal{M}^{mes}$, but can also be used to calculate geometric properties of the Sasaki-Einstein base manifold of the toric Calabi-Yau 4-fold given by $\mathcal{M}^{mes}$. 
One such geometric feature of the Sasaki-Einstein base manifold is its minimum volume.
Given that under mass deformation, the mesonic moduli space $\mathcal{M}^{mes}$ changes, we expect the minimum volume of the corresponding Sasaki-Einstein 7-manifold to change as well. 

The minimum volume of the Sasaki-Einstein base manifold can be directly computed from the Hilbert series of the toric Calabi-Yau 4-fold associated to the brane brick model. 
We summarize the computation in the following paragraph with comments on how the minimum volume changes under mass deformations. 

\paragraph{Minimum Volume of Sasaki-Einstein 7-Manifolds.}
The volume function of the Sasaki-Einstein base $Y_7$ can be directly obtained from the Hilbert series of $\mathcal{M}^{mes} = \mathcal{C}(Y_7)$ as follows, 
\beal{es14a02}
V(b_\mu; Y_7) = \lim_{\beta \rightarrow 0 }
g(t_\mu = \exp [- \beta b_\mu]; \mathcal{C}(Y_7))
~,~
\eea
where $b_\mu$ are the Reeb vector components corresponding to the GLSM fields $p_\mu$.
The limit picks the leading order in $\beta$ in the $\beta$-expansion of the Hilbert series, which was shown in \cite{Martelli:2005tp,Martelli:2006yb} to be directly related to the volume of the Sasaki-Einstein base $Y_7$. 
After obtaining the volume function using \eref{es14a02}, the function can be minimized to a global minimum,
\beal{es14a03}
V_{min} = \min_{b_\mu} V(b_\mu; Y_7) ~.~
\eea

In the case of $4d$ $\mathcal{N}=1$ supersymmetric gauge theories given by brane tilings, the mesonic moduli space $\mathcal{M}^{mes}$ of these theories is a toric Calabi-Yau 3-fold with a Sasaki-Einstein 5-manifold as its base. 
In this context, the minimum volume $V_{min}$ of the Sasaki-Einstein 5-manifold is inversely proportional to the maximized central charge $a_{max} \sim 1/V_{min}$ of the corresponding $4d$ superconformal field theory by the AdS/CFT correspondence \cite{Maldacena:1997re, Morrison:1998cs,Acharya:1998db}.
In the context of brane brick models, the field theoretic interpretation of the minimum volume $V_{min}$ of the Sasaki-Einstein 7-manifolds is less understood. 

Let us summarize what we will observe regarding the effect mass deformation on the minimum volume of the Sasaki-Einstein 7-manifold. As a geometric invariant associated to the theory, it is natural to investigate its behavior under mass deformations, which we expect lead to a decrease in the number of degrees of freedom. 
The geometric counterpart of the decrease in the number of degrees of freedom is well understood for mass deformations of $4d$ $\mathcal{N}=1$ gauge theories corresponding to brane tilings and toric Calabi-Yau 3-folds \cite{Bianchi:2014qma}, where the minimum volume of the associated Sasaki-Einstein 5-manifold increases. We will note that a similar in the examples of brane brick models that we will study, in which the minimum volume of the Sasaki-Einstein 7-manifold increases under mass deformation such that
\beal{es14a05}
\frac{V_{min} }{V_{min}^\prime} < 1 ~,~
\eea
where $V_{min}$ and $V_{min}^\prime$ are the minimum volumes before and after the mass deformation, respectively. Explicit examples of this behavior will be presented in the next section.
\\

\section{Examples \label{sec:4}}

In this section we present various explicit examples of mass deformations connecting toric theories. The general strategy can be summarized as follows. We start from a geometry that admits mass deformations. By this we mean one whose toric diagram contains edges with more than two collinear points and, furthermore, for which deforming that edge as discussed in Section \sref{sec:3} results in another convex toric diagram. Next, we find a toric phase for the parent geometry that contains chiral-Fermi pairs extended between the same pairs of nodes. Those fields are the ones that can potentially become massive. We then search for a toric phase of the target geometry whose quiver differs from the parent one by removal of (some of) the chiral-Fermi pairs. Finally, we determine whether there indeed exists a mass deformation that also connects the $J$- and $E$-terms of both theories.

\subsection{Mass Deformation of the $\cC_{++}$ Theory \label{sec:41}}

\begin{figure}[H]
\begin{center}
\resizebox{0.3\hsize}{!}{
\includegraphics[height=6cm]{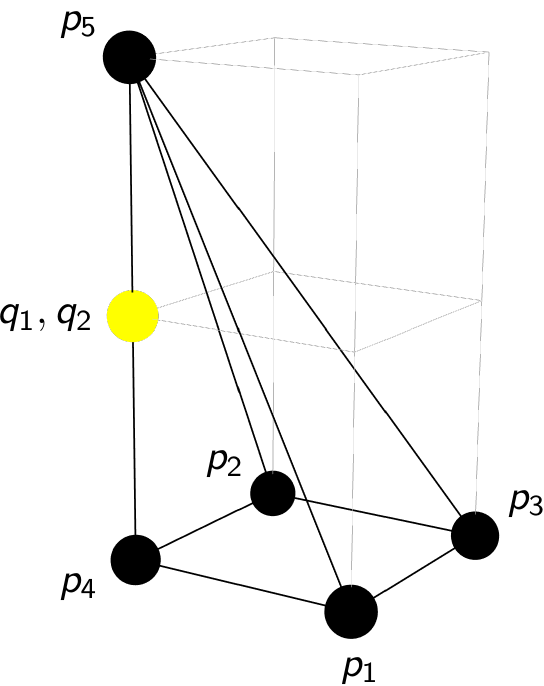} 
}
\caption{
The toric diagram for $\cC_{++}$.
\label{f_toric_1a}}
 \end{center}
 \end{figure}

Let us consider the $\cC_{++}$ geometry, whose toric diagram is shown in \fref{f_toric_1a}. We denote the geometry $\cC_{++}$ following the notation introduced in \cite{Franco:2016fxm} in the context of orbifold reduction. According to this notation, this CY$_4$ toric diagram is obtained by starting from the one for the conifold. The two $+$ subindices indicate that we add two images of one of its points in the same direction transverse to the original $2d$ toric diagram.

\begin{figure}[H]
\begin{center}
\resizebox{0.3\hsize}{!}{
\includegraphics[height=6cm]{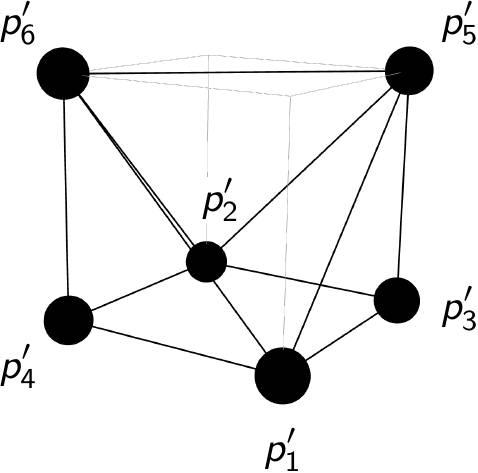} 
}
\caption{
The toric diagram of $H_4$ obtained from a mass deformation of the $\cC_{++}$ theory.
\label{f_toric_1b}}
 \end{center}
 \end{figure}

This toric diagram is a promising starting point, since it exhibits an edge with three collinear points. Furthermore, deforming this edge leads to another convex toric diagram, which is shown in \fref{f_toric_1b}.

\begin{figure}[H]
\begin{center}
\resizebox{0.3\hsize}{!}{
\includegraphics[height=6cm]{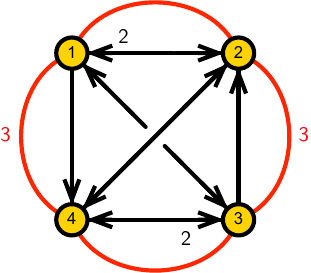} 
}
\caption{
The quiver of the $\cC_{++}$ theory.
\label{f_quiver_1a}}
 \end{center}
 \end{figure}

A possible way of obtaining the gauge theory associated to this geometry (or, equivalently, its brane brick model) is via orbifold reduction starting from the $4d$ $\mathcal{N}=1$ conifold theory \cite{Franco:2016fxm}.\footnote{In general, this would be one of many toric phases.} \fref{f_quiver_1a} shows the quiver diagram for this theory. We observe that this theory contains chiral and Fermi fields stretching between the same nodes, which can become massive in different combinations.

The $J$- and $E$-terms of the $\cC_{++}$ theory take the following form,
\beal{es20a01}
\begin{array}{rcccc}
& & J &  & E  \\
 \Lambda_{12} : & \ \ \  & X_{21} \cdot Y_{12} \cdot Y_{21} -  Y_{21} \cdot Y_{12} \cdot X_{21} &  \ \ \ \  & Z_{13} \cdot X_{32} - X_{14} \cdot Z_{42} \\ 
  \Lambda_{34} : & \ \ \  & X_{43} \cdot Y_{34} \cdot Y_{43} - Y_{43} \cdot Y_{34} \cdot X_{43} &  \ \ \ \  &  Z_{31}  \cdot X_{14} - X_{32} \cdot Z_{24} \\
 \Lambda_{14} : & \ \ \  & Y_{43} \cdot X_{32} \cdot X_{21} - X_{43} \cdot X_{32} \cdot Y_{21} &  \ \ \ \  & Z_{13} \cdot Y_{34} - Y_{12} \cdot Z_{24} \\
 \Lambda_{32} : & \ \ \  & Y_{21} \cdot X_{14} \cdot X_{43} - X_{21} \cdot X_{14} \cdot Y_{43}  &  \ \ \ \  & Z_{31} \cdot Y_{12} - Y_{34} \cdot Z_{42} \\
 \Lambda_{23}^{1} : & \ \ \  & Y_{34} \cdot Y_{43} \cdot X_{32} - X_{32} \cdot Y_{21} \cdot Y_{12} &  \ \ \ \  & Z_{24} \cdot X_{43} - X_{21} \cdot Z_{13} \\ 
 \Lambda_{23}^{2} : & \ \ \  & X_{32} \cdot X_{21} \cdot Y_{12} - Y_{34} \cdot X_{43} \cdot X_{32} &  \ \ \ \  & Z_{24} \cdot Y_{43} - Y_{21} \cdot Z_{13} \\
 \Lambda_{41}^{1} : & \ \ \  & Y_{12} \cdot Y_{21} \cdot X_{14} - X_{14} \cdot Y_{43} \cdot Y_{34} &  \ \ \ \  & Z_{42} \cdot X_{21} - X_{43} \cdot Z_{31} \\
 \Lambda_{41}^{2} : & \ \ \  & X_{14} \cdot X_{43} \cdot Y_{34} - Y_{12} \cdot X_{21} \cdot X_{14} &  \ \ \ \  & Z_{42} \cdot Y_{21} - Y_{43} \cdot Z_{31} 
 \end{array} 
 ~.~
\eea

Using the forward algorithm \cite{Franco:2015tna}, we obtain the $P$-matrix relating fields in the quiver to GLSM fields
\beal{es20a02}
P_\Lambda=
\resizebox{0.3\textwidth}{!}{$
\left(
\begin{array}{c|cc>{\columncolor{Blue!20}}cc>{\columncolor{Orange!20}}c|cc}
\; & p_1 & p_2 & p_3 & p_4 & p_5 & q_1 & q_2\\
\hline
Z_{13} & 0 & 0 & 0 & 0 & 1 & 0 & 1 \\
Z_{24} & 0 & 0 & 0 & 0 & 1 & 0 & 1 \\
Z_{31} & 0 & 0 & 0 & 0 & 1 & 1 & 0 \\
Z_{42} & 0 & 0 & 0 & 0 & 1 & 1 & 0 \\
X_{14} & 0 & 0 & 0 & 1 & 0 & 0 & 1 \\
X_{32} & 0 & 0 & 0 & 1 & 0 & 1 & 0 \\
X_{21} & 0 & 1 & 0 & 0 & 0 & 0 & 0 \\
X_{43} & 0 & 1 & 0 & 0 & 0 & 0 & 0 \\
Y_{12} & 0 & 0 & 1 & 0 & 0 & 0 & 0 \\
Y_{21} & 1 & 0 & 0 & 0 & 0 & 0 & 0 \\
Y_{34} & 0 & 0 & 1 & 0 & 0 & 0 & 0 \\
Y_{43} & 1 & 0 & 0 & 0 & 0 & 0 & 0 \\
\hline
\Lambda_{12} & 0 & 0 & 0 & 1 & 1 & 1 & 1  \\
\Lambda_{34} & 0 & 0 & 0 & 1 & 1 & 1 & 1  \\
\Lambda_{14} & 0 & 0 & 1 & 0 & 1 & 0 & 1  \\ 
\Lambda_{32} & 0 & 0 & 1 & 0 & 1 & 1 & 0  \\
\Lambda^1_{23} & 0 & 1 & 0 & 0 & 1 & 0 & 1  \\
\Lambda^2_{23} & 1 & 0 & 0 & 0 & 1 & 0 & 1  \\ 
\Lambda^1_{41} & 0 & 1 & 0 & 0 & 1 & 1 & 0  \\
\Lambda^2_{41} & 1 & 0 & 0 & 0 & 1 & 1 & 0 
\end{array}
\right)
$}
~,~
\eea
where we have left out the two GLSM fields for simplicity. Using the $P$-matrix, we obtain the $Q_{JE}$ and $Q_D$ matrices that encode the $U(1)$ charges of the GLSM fields due to the $J$- and $E$-terms as well as the $D$-terms of the brane brick model, 
\beal{es20a05}
Q_{JE} = 
\resizebox{0.28\textwidth}{!}{$
\left(
\begin{array}{ccccc|cc}
 p_1 & p_2 & p_3 & p_4 & p_5 & q_1 & q_2 
\\
\hline
1 & 1 & 1 & 0 & 1 & 0 & 0 \\
1 & 1 & 1 & -1 & 0 & 1 & 1 \\
\end{array}
\right) 
$}
~~,~~
Q_{D} = 
\resizebox{0.3\textwidth}{!}{$
\left(
\begin{array}{ccccc|cc}
 p_1 & p_2 & p_3 & p_4 & p_5 & q_1 & q_2 
\\
\hline
-1 & -1 & 0 & 1 & 0 & -1 & 0  \\ 
0 & 0 & -1 & 0 & 0 & -1 & 0 \\ 
0 & 0 & 1 & 0 & 0 & 1 & 0 
\end{array}
\right) 
$}
~.~
\eea

The toric diagram is encoded in the $G_t$-matrix, which is the kernel of the concatenation of the $Q_{JE}$ and $Q_{D}$ matrices. Each of its columns provides the coordinates of the GLSM fields in the toric diagram. In this case, we get
\beal{es20a07}
G_{t} = 
\resizebox{0.28\textwidth}{!}{$
\left(
\begin{array}{ccccc|cc}
 p_1 & p_2 & p_3 & p_4 & p_5 & q_1 & q_2 
\\
\hline
 1 & 1 & 1 & 1 & 1 & 1 & 1  \\
 1 & 0 & 1 & 0 & 0 & 0 & 0  \\
 0 & 1 & 1 & 0 & 0 & 0 & 0  \\
 0 & 0 & 0 & 0 & 2 & 1 & 1  \\
\end{array}
\right)
$}
~,~
\eea
which indeed agrees with the desired toric diagram of $\cC_{++}$, shown in \fref{f_toric_1a}.

Following the $P$-matrix in \eref{es20a02}, we can assign global symmetry charges $\alpha_\mu$ corresponding to the extremal brick matchings to chiral and Fermi fields as shown in \tref{t_global_1a}.
We recall from section \sref{sec:22} that the sum of the charge vectors $\alpha_\mu$ vanishes,
\beal{es20a06b}
\alpha_1 + \alpha_2 + \alpha_3 + \alpha_4 + \alpha_5 = 0 ~,~
\eea
for the $J$- and $E$-terms to be invariant under the global symmetry.

\begin{table}[H]
\begin{center}
\begin{tabular}{|c|p{1.3cm}|}
\hline
$Z_{13}$ & $\alpha_5$\\
$Z_{24}$ & $\alpha_5$\\
$Z_{31}$ & $\alpha_5$\\
$Z_{42}$ & $\alpha_5$\\
\hline
\end{tabular}
~
\begin{tabular}{|c|p{1.3cm}|}
\hline
$X_{14}$ & $\alpha_4$\\
$X_{32}$ & $\alpha_4$\\
$X_{21}$ & $\alpha_2$\\
$X_{43}$ & $\alpha_2$\\
\hline
\end{tabular}
~
\begin{tabular}{|c|p{1.3cm}|}
\hline
$Y_{12}$ & $\alpha_3$\\
$Y_{21}$ & $\alpha_1$\\
$Y_{34}$ & $\alpha_3$\\
$Y_{43}$ & $\alpha_1$\\
\hline
\end{tabular}
~
\begin{tabular}{|c|p{1.3cm}|}
\hline
$\Lambda_{12}$ & $\alpha_4 +\alpha_5$\\
$\Lambda_{34}$ & $\alpha_4 +\alpha_5$\\
$\Lambda_{14}$ & $\alpha_3 +\alpha_5$\\
$\Lambda_{32}$ & $\alpha_3 +\alpha_5$\\
\hline
\end{tabular}
~
\begin{tabular}{|c|p{1.3cm}|}
\hline
$\Lambda_{23}^1$ & $\alpha_2 +\alpha_5$\\
$\Lambda_{23}^2$ & $\alpha_1 +\alpha_5$\\
$\Lambda_{41}^1$ & $\alpha_2 +\alpha_5$\\
$\Lambda_{41}^2$ & $\alpha_1 +\alpha_5$\\
\hline
\end{tabular}
\caption{
Global symmetry charge assignments for chiral and Fermi fields of the $\cC_{++}$ theory.
\label{t_global_1a}}
 \end{center}
 \end{table}

Let us consider the following mass deformation of the $\cC_{++}$ theory, corresponding to adding the terms shown in blue to the $E$-terms,
\beal{es20a10}
\begin{array}{rrrclcrcl}
& & & J   & & & E & \textcolor{blue}{+}  & \textcolor{blue}{\Delta E} \\
 \Lambda_{12} : & \ \ \  & X_{21} \cdot Y_{12} \cdot Y_{21}  &-&   Y_{21} \cdot Y_{12} \cdot X_{21}  & \ \ \ \  & \textcolor{blue}{ - Y_{12}}+ Z_{13} \cdot X_{32}  &-&  X_{14} \cdot Z_{42} \\ 
 \Lambda_{34} : & \ \ \  & X_{43} \cdot Y_{34} \cdot Y_{43}  &-&  Y_{43} \cdot Y_{34} \cdot X_{43}  & \ \ \ \  &  \textcolor{blue}{+Y_{34}}+Z_{31} \cdot X_{14}  &-&  X_{32} \cdot Z_{24}  \\
 \end{array} 
 ~.~
\eea
When the massive Fermi fields are integrated out, 
the two massive chiral fields $Y_{12}$ and $Y_{34}$ undergo a field replacement as follows,
\beal{es20a11}
Y_{12} & = + Z_{13} \cdot X_{32} - X_{14} \cdot Z_{42} ~,~\nn\\
 Y_{34} & = - Z_{31} \cdot X_{14} + X_{32} \cdot Z_{24} ~.~
\eea
We note that the massive chiral fields $Y_{12}, Y_{34} $ and the massive Fermi fields $\bar{\Lambda}_{12},\bar{\Lambda}_{34}$ are all part of brick matching $p_3$, which is highlighted in blue in the $P$-matrix in \eref{es20a02}.

Applying the chiral field replacement in \eref{es20a11}, we obtain
\beal{es20a12}
\resizebox{0.95\textwidth}{!}{$
\begin{array}{rcl}
 \Lambda_{14} : & J:  & X_{43} \cdot X_{32} \cdot Y_{21} - Y_{43} \cdot X_{32} \cdot X_{21}   \\
 & E: & Z_{13} \cdot Z_{31} \cdot X_{14} - \textcolor{BrickRed}{Z_{13} \cdot X_{32} \cdot Z_{24}}  + \textcolor{BrickRed}{Z_{13} \cdot X_{32} \cdot Z_{24}} - X_{14} \cdot Z_{42} \cdot Z_{24}    \\[.1cm]
 \Lambda_{32} : & J:  & X_{21} \cdot X_{14} \cdot Y_{43} - Y_{21} \cdot X_{14} \cdot X_{43}    \\
  & E: & Z_{31} \cdot Z_{13} \cdot X_{32} - \textcolor{BrickRed}{Z_{31} \cdot X_{14} \cdot Z_{42} + Z_{31} \cdot X_{14} \cdot Z_{42}} - X_{32} \cdot Z_{24} \cdot Z_{42}   \\[.1cm] 
 \Lambda_{23}^{1} : & J:  & \textcolor{ForestGreen}{X_{32} \cdot Y_{21} \cdot Z_{13} \cdot X_{32}} - X_{32} \cdot Y_{21}\cdot X_{14} \cdot Z_{42} + Z_{31} \cdot X_{14} \cdot Y_{43} \cdot X_{32} -  \textcolor{ForestGreen}{X_{32} \cdot Z_{24} \cdot Y_{43} \cdot X_{32}}    \\
& E: & Z_{24} \cdot X_{43} - X_{21} \cdot Z_{13}    \\[.1cm]
 \Lambda_{23}^{2} : & J:  & \textcolor{ForestGreen}{ X_{32} \cdot X_{21} \cdot Z_{13} \cdot X_{32} } - X_{32} \cdot X_{21} \cdot X_{14} \cdot Z_{42} + Z_{31} \cdot X_{14} \cdot X_{43} \cdot X_{32} - \textcolor{ForestGreen}{ X_{32} \cdot Z_{24} \cdot X_{43} \cdot X_{32}}    \\
 & E: & Z_{24} \cdot Y_{43} - Y_{21} \cdot Z_{13}   \\[.1cm]
 \Lambda_{41}^{1} : & J: & \textcolor{ForestGreen}{-X_{14} \cdot Y_{43} \cdot Z_{31} \cdot X_{14}} + X_{14} \cdot Y_{43} \cdot X_{32} \cdot Z_{24} - Z_{13} \cdot X_{32} \cdot Y_{21} \cdot X_{14} \textcolor{ForestGreen}{+ X_{14} \cdot Z_{42} \cdot Y_{21} \cdot X_{14}}   \\
 & E: & Z_{42} \cdot X_{21} - X_{43} \cdot Z_{31}    \\[.1cm] 
 \Lambda_{41}^{2} : & J:  & \textcolor{ForestGreen}{- X_{14} \cdot X_{43} \cdot Z_{31} \cdot X_{14}} + X_{14} \cdot X_{43} \cdot X_{32} \cdot Z_{24} - Z_{13} \cdot X_{32} \cdot X_{21} \cdot X_{14} \textcolor{ForestGreen}{+ X_{14} \cdot Z_{42} \cdot X_{21} \cdot X_{14}}   \\
 & E: & Z_{42} \cdot Y_{21} - Y_{43} \cdot Z_{31}    \\[.1cm] 
 \end{array}
$}
~.~
\nn\\
\eea
Above, we note that the red terms cancel each other trivially, while the green terms cancel due to the $E$-terms, whereas the red terms cancel each other trivially.

\begin{figure}[H]
\begin{center}
\resizebox{0.3\hsize}{!}{
\includegraphics[height=6cm]{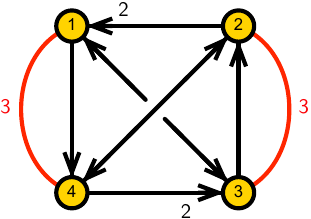} 
}
\caption{
The quiver obtained from the $\cC_{++}$ theory after the mass deformation in \eqref{es20a10}.
\label{f_quiver_1b}}
 \end{center}
 \end{figure}

\fref{f_quiver_1b} shows the resulting quiver.

The corresponding $J$- and $E$-terms are
\beal{es20a20}
\begin{array}{rrrclcrcl}
& & & J  & &  & & E   \\
 \Lambda_{32} : & \ \ \  & X_{21} \cdot X_{14} \cdot Y_{43}  &-&  Y_{21} \cdot X_{14} \cdot X_{43}   & \ \ \ \  & Z_{31} \cdot Z_{13} \cdot X_{32} &-& X_{32}\cdot Z_{24} \cdot Z_{42}   \\ 
 \Lambda_{14} : & \ \ \  & X_{43} \cdot X_{32} \cdot Y_{21}  &-&  Y_{43} \cdot X_{32} \cdot X_{21}   & \ \ \ \  & Z_{13} \cdot Z_{31}\cdot X_{14}  &-&  X_{14} \cdot Z_{42}\cdot Z_{24}   \\
 \Lambda_{23}^{1} : & \ \ \  & Z_{31} \cdot X_{14} \cdot Y_{43} \cdot X_{32}  &-&  X_{32} \cdot Y_{21}\cdot X_{14} \cdot Z_{42}   & \ \ \ \  & Z_{24} \cdot X_{43}  &-&  X_{21} \cdot Z_{13}   \\ 
 \Lambda_{23}^{2} : & \ \ \  & Z_{31} \cdot X_{14} \cdot X_{43} \cdot X_{32}  &-&  X_{32} \cdot X_{21} \cdot X_{14} \cdot Z_{42}    & \ \ \ \  & Z_{24} \cdot Y_{43}  &-&  Y_{21} \cdot Z_{13}   \\ 
 \Lambda_{41}^{1} : & \ \ \  & X_{14} \cdot Y_{43} \cdot X_{32} \cdot Z_{24}  &-&  Z_{13} \cdot X_{32} \cdot Y_{21} \cdot X_{14}   & \ \ \ \  & Z_{42} \cdot X_{21}  &-&  X_{43} \cdot Z_{31}   \\
 \Lambda_{41}^{2} : & \ \ \  & X_{14} \cdot X_{43} \cdot X_{32} \cdot Z_{24}  &-&  Z_{13} \cdot X_{32} \cdot X_{21} \cdot X_{14}   & \ \ \ \  & Z_{42} \cdot Y_{21}  &-&  Y_{43} \cdot Z_{31}   \\
 \end{array} 
 ~,~
 \nn\\
\eea
This gauge theory is indeed one of the toric phases of $H_4$ which was first studied in \cite{Franco:2017cjj,Franco:2018qsc}. Below we confirm this by direct application of the forward algorithm.

From the $J$- and $E$-terms of the new $H_4$ theory, we obtain the corresponding $P$-matrix,
\beal{es20a21}
P_{\Lambda} = 
\resizebox{0.25\textwidth}{!}{$
\left(
\begin{array}{c|cccccc}
 & p'_1 & p'_2 & p'_3 & p'_4 & p'_5 & p'_6 \\
\hline
Z_{13} & 0 & 0 & 0 & 0 & 1 & 0  \\ 
Z_{24} & 0 & 0 & 0 & 0 & 1 & 0 \\
Z_{31} & 0 & 0 & 0 & 0 & 0 & 1 \\ 
Z_{42} & 0 & 0 & 0 & 0 & 0 & 1 \\ 
X_{14} & 0 & 0 & 0 & 1 & 0 & 0 \\ 
X_{21} & 1 & 0 & 0 & 0 & 0 & 0 \\
X_{32} & 0 & 0 & 1 & 0 & 0 & 0 \\ 
X_{43} & 1 & 0 & 0 & 0 & 0 & 0 \\ 
Y_{21} & 0 & 1 & 0 & 0 & 0 & 0 \\ 
Y_{43} & 0 & 1 & 0 & 0 & 0 & 0 \\ 
\hline
\Lambda_{32} & 0 & 0 & 1 & 0 & 1 & 1 \\ 
\Lambda_{14} & 0 & 0 & 0 & 1 & 1 & 1 \\ 
\Lambda^1_{23} & 1 & 0 & 0 & 0 & 1 & 0 \\ 
\Lambda^2_{23} & 0 & 1 & 0 & 0 & 1 & 0  \\ 
\Lambda^1_{41} & 1 & 0 & 0 & 0 & 0 & 1  \\ 
\Lambda^2_{41} & 0 & 1 & 0 & 0 & 0 & 1 
\end{array}
\right) 
$}
~,~
\eea
where we have omitted the two extra GLSM fields.
Note that the brick matchings have now been relabeled as $p'_\mu$ in order to distinguish them with the brick matchings $p_\mu$ from the original $\cC_{++}$ theory.
Using the $P$-matrix, we identify the charge matrices under the $J$- and $E$-terms as well as the $D$-terms,
\beal{es20a22}
Q_{JE} = 
\resizebox{0.23\textwidth}{!}{$
\left(
\begin{array}{cccccc}
p'_1 & p'_2 & p'_3 & p'_4 & p'_5 & p'_6 
\\
\hline
1 & 1 & 0 & 0 & 1 & 1
\end{array}
\right) 
$}
~,~
Q_{D} = 
\resizebox{0.25\textwidth}{!}{$
\left(
\begin{array}{cccccc}
p'_1 & p'_2 & p'_3 & p'_4 & p'_5 & p'_6 
\\
\hline
0 & 0 & 0 & 1 & 1 & 0  \\ 
0 & 0 & -1 & 0 & 0 & -1 \\ 
0 & 0 & 1 & 0 & 0 & 1
\end{array}
\right) 
$}
~.~
\eea
The resulting toric diagram for the $H_4$ theory is given by
\beal{es20a23}
G_{t} = 
\resizebox{0.23\textwidth}{!}{$
\left(
\begin{array}{cccccc}
 p'_1 & p'_2 & p'_3 & p'_4 & p'_5 & p'_6 
\\
\hline
1 & 1 & 1 & 1 & 1 & 1  \\
1 & 0 & 1 & 0 & 1 & 0 \\
0 & 1 & 1 & 0 & 1 & 0  \\
 0 & 0 & 0 & 0 & 1 & 1 
\end{array}
\right) 
$}
~,~
\eea
and illustrated in \fref{f_toric_1b}.

When we compare the toric Calabi-Yau 4-fold before and after the mass deformation, we note that most of the points keep the same coordinates (up to at most $SL(3,\mathbb{Z}) transformations)$. 
Only the toric point corresponding to extremal brick matching $p_5$ highlighted in orange in \eref{es20a02} changes its coordinates.
By doing so, it turns the non-extremal brick matchings $q_1,q_2$ into a single new extremal brick matching $p'_6$.

\begin{figure}[H]
\begin{center}
\resizebox{0.7\hsize}{!}{
\includegraphics[height=6cm]{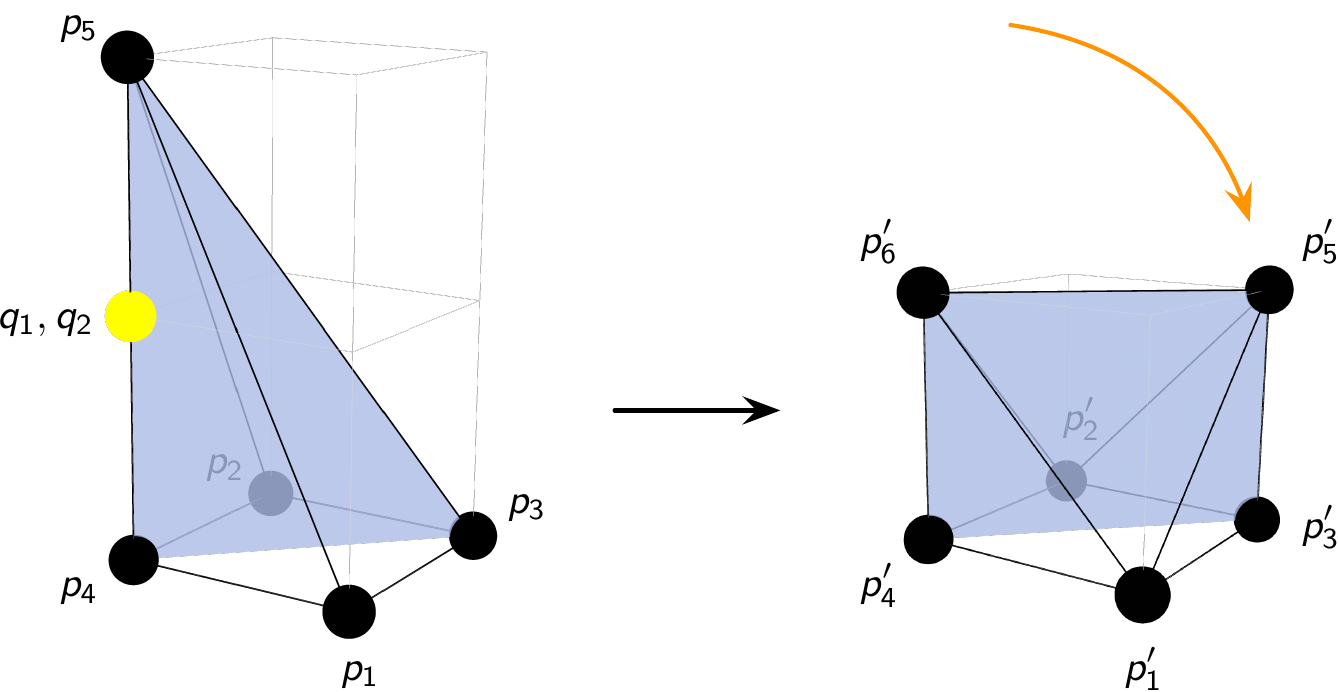} 
}
\caption{
The toric diagram of $\cC_{++}$ changing into the toric diagram of $H_4$ under mass deformation. Brick matchings $p_3$, $p_4$ and $p_5$ define a 2-dimensional plane (blue) and correspond to the massive terms added to the $J$- and $E$-terms. On this plane, the point associated to $p_5$ moves to a new position due to the mass deformation.
\label{f_toric_1a3}}
 \end{center}
 \end{figure}

As illustrated in \fref{f_toric_1a3}, the massive brick matching $p_3$, the unaffected brick matching $p_4$ and the moving brick matching $p_5$ form a plane on which the transformation of the toric diagram takes place. 
After the mass deformation, brick matchings $p_3 \mapsto p'_3$, $p_4 \mapsto p'_4$, $p_5 \mapsto p'_5$ and the new extremal brick matching $q_1,q_2 \mapsto p'_6$ are all on the same plane highlighted in blue in \fref{f_toric_1a3}.
The toric point corresponding to $p_5$ moves along this plane to its new position under mass deformation.

When we consider the global symmetry charges on chiral and Fermi fields summarized in \tref{t_global_1a}, the mass terms added in \eref{es20a10}, which correspond to the plaquettes
\beal{es20a25}
\bar{\Lambda}_{12} \cdot Y_{12} ~,~
\bar{\Lambda}_{34} \cdot Y_{34} ~,~
\eea
carry the following charges,
\beal{es20a26}
\alpha(\bar{\Lambda}_{12} \cdot Y_{12} ) = \alpha_3 - \alpha_4 - \alpha_5 ~,~
\alpha(\bar{\Lambda}_{34} \cdot Y_{34} ) = \alpha_3 - \alpha_4 - \alpha_5 ~.~
\eea
We note that massive plaquettes carry the same non-zero global symmetry charges. 
These non-zero global symmetry charges exactly correspond to brick matchings $p_3$, $p_4$ and $p_5$ whose corresponding points in the toric diagram are on the blue plane in \fref{f_toric_1a3}.
Given that the non-zero global symmetry charges in \eref{es20a26}, we can identify the blue plane in the toric diagrams in \fref{f_toric_1a3} as the plane along which part of the global symmetry of the $\cC_{++}$ theory breaks under mass deformation. 
The global symmetry of the resulting $H_4$ contain linear combinations of the original symmetries under which the massive plaquettes are neutral. 

When we compute the minimum volumes of the Sasaki-Einstein 7-manifolds corresponding to the toric Calabi-Yau 4-folds before and after the mass deformation, we obtain respectively the following,
\beal{es20a30}
V_{min}^{\cC_{++}}  = \frac{8}{27} \simeq 0.296296 ~,~
V_{min}^{H_4} = \frac{11 + 5 \sqrt{5}}{64}  \simeq 0.346563 ~.~
\eea
The ratio between the two volumes is
\beal{es20a31}
\frac{V_{min}^{\cC_{++}}}{V_{min}^{H_4}} \simeq 0.854945 < 1 ~,~
\eea
indicating that the minimum volume of the Sasaki-Einstein 7-manifolds increases with mass deformation, as in the CY$_3$ case. 

The mass deformation of the $\cC_{++}$ theory discussed in this section
relates to the Klebanov-Witten flow between the $4d$ $\mathcal{N}=1$ theories associated to the 
$\mathbb{C}^2 / \mathbb{Z}_2 \times \mathbb{C} $ orbifold and the conifold 
by orbifold reduction \cite{Franco:2016fxm}. 
The 2-dimensional toric diagrams for $\mathbb{C}^2 / \mathbb{Z}_2 \times \mathbb{C}$ and $\cC$ are the 2-dimensional intersections between the plane along which part of the global symmetry of the mass deformed brane brick model breaks and the 3-dimensional toric diagrams for $\cC_{++}$ and $H_4$ as highlighted in blue in \fref{f_toric_1a3}.
\\

\subsection{Mass Deformation of the $\mathbb{F}_{0,+-}$ Theory \label{sec:42}}

Let us now consider $\mathbb{F}_{0,+-}$, whose toric diagram is shown in \fref{f_toric_1a}. Once again, we use the notation introduce in \cite{Franco:2016fxm}, which indicates that this toric diagram is obtained by starting from $F_0$ and adding two images of one of its points, in this case a corner, one above and the other one below the original plane. Like the previous example, this toric diagram contains an edge with three collinear points.

\begin{figure}[H]
\begin{center}
\resizebox{0.4\hsize}{!}{
\includegraphics[height=6cm]{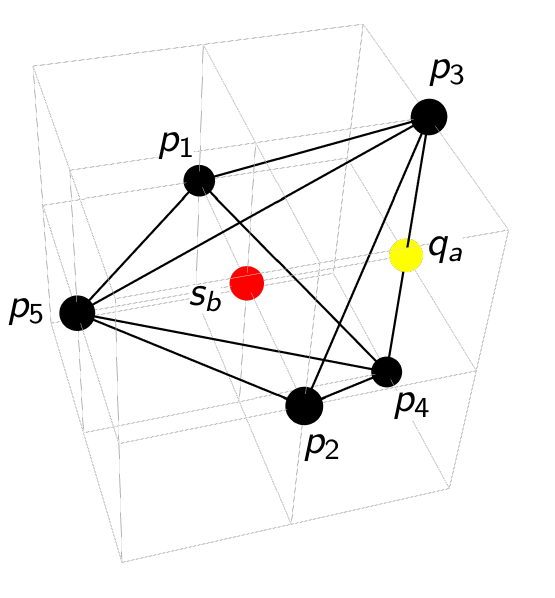} 
}
\caption{
The toric diagram for $\mathbb{F}_{0,+-}$. 
\label{f_toric_2a}}
 \end{center}
 \end{figure}

\begin{figure}[H]
\begin{center}
\resizebox{0.35\hsize}{!}{
\includegraphics[height=6cm]{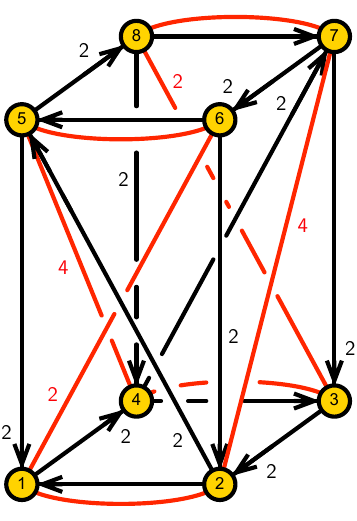} 
}
\caption{
The quiver of the $\mathbb{F}_{0,+-}$ theory.
\label{f_quiver_2a}}
 \end{center}
 \end{figure}

We construct one of the $2d$ $(0,2)$ phases for this geometry via orbifold reduction starting from a $4d$ $\mathcal{N}=1$ theory for $\mathbb{F}_0$ \cite{Feng:2001xr,Franco:2005rj}.\footnote{The $2d$ $(0,2)$ gauge theory for this geometry can alternatively be constructed by orbifold reduction starting from the $4d$ $\mathcal{N}=1$ theory  for a $\mathbb{C}^3/\mathbb{Z}_{4}$ orbifold \cite{Franco:2016fxm}.} More specifically, we use the first phase of $\mathbb{F}_0$ as starting point. \fref{f_quiver_2a} shows the resulting quiver, which contains four chiral-Fermi pairs that can become massive.

The $J$- and $E$-terms for this theory are
\beal{es30a01}
\begin{array}{rrrclcrcl}
& & & J  & & &  & E   \\
\Lambda_{21} : & \ & Y_{14} \cdot Y_{43} \cdot X_{32}  &-&  X_{14} \cdot Y_{43} \cdot Y_{32}   & \ \ \ \ &  U_{25} \cdot P_{51}  &-&  W_{25} \cdot Q_{51}  \\
\Lambda_{43} : & \ & X_{32} \cdot Y_{21} \cdot Y_{14}  &-&  Y_{32} \cdot Y_{21} \cdot X_{14}   & \ \ \ \ &  W_{47} \cdot  Q_{73}  &-&  U_{47} \cdot P_{73}  \\
\Lambda_{65} : & \ & Y_{58} \cdot Y_{87} \cdot X_{76}  &-&  X_{58} \cdot Y_{87} \cdot Y_{76}   & \ \ \ \ & Q_{62} \cdot  W_{25}  &-&  P_{62} \cdot U_{25}   \\
 \Lambda_{87} : & \ & X_{76} \cdot Y_{65} \cdot Y_{58}  &-&  Y_{76} \cdot Y_{65} \cdot X_{58}   & \ \ \ \ &  P_{84} \cdot U_{47}  &-&  Q_{84} \cdot  W_{47}   \\
 \Lambda^1_{54} : & \ & W_{47} \cdot X_{76} \cdot Y_{65}  &-&  Y_{43} \cdot X_{32} \cdot W_{25}   & \ \ \ \ &  Y_{58} \cdot  Q_{84}  &-&  Q_{51} \cdot Y_{14}  \\
 \Lambda^2_{54} : & \ & W_{47} \cdot Y_{76} \cdot Y_{65}  &-&  Y_{43} \cdot Y_{32} \cdot W_{25}   & \ \ \ \ &  Q_{51} \cdot X_{14}  &-&  X_{58} \cdot  Q_{84}   \\
 \Lambda^3_{54} : & \ & U_{47} \cdot Y_{76} \cdot Y_{65}  &-&   Y_{43} \cdot Y_{32} \cdot U_{25}   & \ \ \ \ &  X_{58} \cdot  P_{84}  &-&  P_{51} \cdot X_{14}    \\
 \Lambda^4_{54} : & \ & U_{47} \cdot X_{76} \cdot Y_{65}  &-&   Y_{43} \cdot X_{32} \cdot U_{25}   & \ \ \ \ &  P_{51} \cdot Y_{14}  &-&  Y_{58} \cdot  P_{84}   \\
 \Lambda^1_{72} : & \ & Y_{21} \cdot Y_{14} \cdot W_{47}  &-&   W_{25} \cdot Y_{58} \cdot Y_{87}   & \ \ \ \ &  X_{76} \cdot  Q_{62}  &-&  Q_{73} \cdot X_{32}  \\
 \Lambda^2_{72} : & \ & Y_{21} \cdot Y_{14} \cdot U_{47}  &-&   U_{25} \cdot Y_{58} \cdot Y_{87}   & \ \ \ \ &  P_{73} \cdot X_{32}  &-&  X_{76} \cdot  P_{62}   \\
 \Lambda^3_{72} : & \ & Y_{21} \cdot X_{14} \cdot U_{47}  &-&   U_{25} \cdot X_{58} \cdot Y_{87}   & \ \ \ \ &  Y_{76} \cdot  P_{62}   &-&  P_{73} \cdot Y_{32}   \\
 \Lambda^4_{72} : & \ & Y_{21} \cdot X_{14} \cdot W_{47}  &-&   W_{25} \cdot X_{58} \cdot Y_{87}   & \ \ \ \ &  Q_{73} \cdot Y_{32}  &-&  Y_{76} \cdot  Q_{62}   \\
 \Lambda^1_{61} : & \ &  Y_{14} \cdot W_{47} \cdot X_{76} &-&   X_{14} \cdot W_{47} \cdot Y_{76}   & \ \ \ \ & Y_{65} \cdot  Q_{51}  &-&   Q_{62} \cdot Y_{21}   \\
 \Lambda^2_{61} : & \ & Y_{14} \cdot U_{47} \cdot X_{76} &-&   X_{14} \cdot U_{47} \cdot Y_{76}   & \ \ \ \ &  P_{62} \cdot Y_{21}  &-&  Y_{65} \cdot  P_{51}   \\
\Lambda^1_{83} : & \ &  Y_{32} \cdot W_{25} \cdot X_{58} &-&   X_{32} \cdot W_{25} \cdot Y_{58}   & \ \ \ \ &  Y_{87} \cdot  Q_{73}  &-&  Q_{84} \cdot Y_{43}   \\
 \Lambda^2_{83} : & \ &  Y_{32} \cdot U_{25} \cdot X_{58} &-&   X_{32} \cdot U_{25} \cdot Y_{58}   & \ \ \ \ &  P_{84} \cdot Y_{43}  &-&  Y_{87} \cdot  P_{73}   \\
\end{array}
~.~
\eea

Following the forward algorithm, we obtain the $P$-matrix given by
\beal{es30a02}
P_\Lambda=
\resizebox{0.55\textwidth}{!}{$
\left(
\begin{array}{c|ccc>{\columncolor{Orange!20}}c>{\columncolor{Blue!20}}c|cc|cccccccccc}
 & p_1 & p_2 & p_3 & p_4 & p_5 & q_1 & q_2 & s_1 & s_2 & s_3 & s_4 & s_5 & s_6 & s_7 & s_8 & s_9 & s_{10} \\
\hline
P_{51} & 0 & 0 & 1 & 0 & 0 & 0 & 1 & 0 & 0 & 0 & 0 & 0 & 0 & 0 & 0 & 1 & 1 \\
P_{62} & 0 & 0 & 1 & 0 & 0 & 0 & 1 & 0 & 0 & 0 & 0 & 0 & 0 & 1 & 1 & 0 & 0 \\
P_{73} & 0 & 0 & 1 & 0 & 0 & 0 & 1 & 0 & 0 & 0 & 0 & 0 & 1 & 0 & 1 & 0 & 0 \\
P_{84} & 0 & 0 & 1 & 0 & 0 & 0 & 1 & 0 & 0 & 0 & 0 & 1 & 0 & 0 & 0 & 0 & 1 \\
Q_{51} & 0 & 0 & 0 & 1 & 0 & 0 & 1 & 0 & 0 & 0 & 0 & 0 & 0 & 0 & 0 & 1 & 1 \\
Q_{62} & 0 & 0 & 0 & 1 & 0 & 0 & 1 & 0 & 0 & 0 & 0 & 0 & 0 & 1 & 1 & 0 & 0 \\
Q_{73} & 0 & 0 & 0 & 1 & 0 & 0 & 1 & 0 & 0 & 0 & 0 & 0 & 1 & 0 & 1 & 0 & 0 \\
Q_{84} & 0 & 0 & 0 & 1 & 0 & 0 & 1 & 0 & 0 & 0 & 0 & 1 & 0 & 0 & 0 & 0 & 1 \\
U_{25} & 0 & 0 & 0 & 1 & 0 & 1 & 0 & 0 & 0 & 0 & 1 & 0 & 0 & 0 & 0 & 0 & 0 \\
U_{47} & 0 & 0 & 0 & 1 & 0 & 1 & 0 & 0 & 0 & 1 & 0 & 0 & 0 & 0 & 0 & 0 & 0 \\
W_{25} & 0 & 0 & 1 & 0 & 0 & 1 & 0 & 0 & 0 & 0 & 1 & 0 & 0 & 0 & 0 & 0 & 0 \\
W_{47} & 0 & 0 & 1 & 0 & 0 & 1 & 0 & 0 & 0 & 1 & 0 & 0 & 0 & 0 & 0 & 0 & 0 \\
X_{14} & 0 & 1 & 0 & 0 & 0 & 0 & 0 & 0 & 1 & 0 & 0 & 1 & 0 & 0 & 0 & 0 & 0 \\
X_{32} & 0 & 1 & 0 & 0 & 0 & 0 & 0 & 1 & 0 & 0 & 0 & 0 & 0 & 1 & 0 & 0 & 0 \\
X_{58} & 0 & 1 & 0 & 0 & 0 & 0 & 0 & 0 & 1 & 0 & 0 & 0 & 0 & 0 & 0 & 1 & 0 \\
X_{76} & 0 & 1 & 0 & 0 & 0 & 0 & 0 & 1 & 0 & 0 & 0 & 0 & 1 & 0 & 0 & 0 & 0 \\
Y_{14} & 1 & 0 & 0 & 0 & 0 & 0 & 0 & 0 & 1 & 0 & 0 & 1 & 0 & 0 & 0 & 0 & 0 \\
Y_{21} & 0 & 0 & 0 & 0 & 1 & 0 & 0 & 0 & 0 & 0 & 1 & 0 & 0 & 0 & 0 & 1 & 1 \\
Y_{32} & 1 & 0 & 0 & 0 & 0 & 0 & 0 & 1 & 0 & 0 & 0 & 0 & 0 & 1 & 0 & 0 & 0 \\
Y_{43} & 0 & 0 & 0 & 0 & 1 & 0 & 0 & 0 & 0 & 1 & 0 & 0 & 1 & 0 & 1 & 0 & 0 \\
Y_{58} & 1 & 0 & 0 & 0 & 0 & 0 & 0 & 0 & 1 & 0 & 0 & 0 & 0 & 0 & 0 & 1 & 0 \\
Y_{65} & 0 & 0 & 0 & 0 & 1 & 0 & 0 & 0 & 0 & 0 & 1 & 0 & 0 & 1 & 1 & 0 & 0 \\
Y_{76} & 1 & 0 & 0 & 0 & 0 & 0 & 0 & 1 & 0 & 0 & 0 & 0 & 1 & 0 & 0 & 0 & 0 \\
Y_{87} & 0 & 0 & 0 & 0 & 1 & 0 & 0 & 0 & 0 & 1 & 0 & 1 & 0 & 0 & 0 & 0 & 1 \\
\hline
\Lambda_{21} & 0 & 0 & 1 & 1 & 0 & 1 & 1 & 0 & 0 & 0 & 1 & 0 & 0 & 0 & 0 & 1 & 1 \\
\Lambda_{43} & 0 & 0 & 1 & 1 & 0 & 1 & 1 & 0 & 0 & 1 & 0 & 0 & 1 & 0 & 1 & 0 & 0 \\
\Lambda_{65} & 0 & 0 & 1 & 1 & 0 & 1 & 1 & 0 & 0 & 0 & 1 & 0 & 0 & 1 & 1 & 0 & 0 \\
\Lambda_{87} & 0 & 0 & 1 & 1 & 0 & 1 & 1 & 0 & 0 & 1 & 0 & 1 & 0 & 0 & 0 & 0 & 1 \\
\Lambda^1_{54} & 1 & 0 & 0 & 1 & 0 & 0 & 1 & 0 & 1 & 0 & 0 & 1 & 0 & 0 & 0 & 1 & 1 \\
\Lambda^2_{54} & 0 & 1 & 0 & 1 & 0 & 0 & 1 & 0 & 1 & 0 & 0 & 1 & 0 & 0 & 0 & 1 & 1 \\
\Lambda^3_{54} & 0 & 1 & 1 & 0 & 0 & 0 & 1 & 0 & 1 & 0 & 0 & 1 & 0 & 0 & 0 & 1 & 1 \\
\Lambda^4_{54} & 1 & 0 & 1 & 0 & 0 & 0 & 1 & 0 & 1 & 0 & 0 & 1 & 0 & 0 & 0 & 1 & 1 \\
\Lambda^1_{72} & 0 & 1 & 0 & 1 & 0 & 0 & 1 & 1 & 0 & 0 & 0 & 0 & 1 & 1 & 1 & 0 & 0 \\
\Lambda^2_{72} & 0 & 1 & 1 & 0 & 0 & 0 & 1 & 1 & 0 & 0 & 0 & 0 & 1 & 1 & 1 & 0 & 0 \\
\Lambda^3_{72} & 1 & 0 & 1 & 0 & 0 & 0 & 1 & 1 & 0 & 0 & 0 & 0 & 1 & 1 & 1 & 0 & 0 \\
\Lambda^4_{72} & 1 & 0 & 0 & 1 & 0 & 0 & 1 & 1 & 0 & 0 & 0 & 0 & 1 & 1 & 1 & 0 & 0 \\
\Lambda^1_{61} & 0 & 0 & 0 & 1 & 1 & 0 & 1 & 0 & 0 & 0 & 1 & 0 & 0 & 1 & 1 & 1 & 1 \\
\Lambda^2_{61} & 0 & 0 & 1 & 0 & 1 & 0 & 1 & 0 & 0 & 0 & 1 & 0 & 0 & 1 & 1 & 1 & 1 \\
\Lambda^1_{83} & 0 & 0 & 0 & 1 & 1 & 0 & 1 & 0 & 0 & 1 & 0 & 1 & 1 & 0 & 1 & 0 & 1 \\
\Lambda^2_{83} & 0 & 0 & 1 & 0 & 1 & 0 & 1 & 0 & 0 & 1 & 0 & 1 & 1 & 0 & 1 & 0 & 1 
 \end{array}
\right)
$}
~,~
\eea
where we have left out the $6$ extra GLSM fields. 

The $U(1)$ charges of brick matchings corresponding to the $J$- and $E$-terms as well as the $D$-terms are \beal{es30a05}
Q_{JE} = 
\resizebox{0.58\textwidth}{!}{$
\left(
\begin{array}{ccccc|cc|cccccccccc}
 p_1 & p_2 & p_3 & p_4 & p_5 & q_1 & q_2 & s_1 & s_2 & s_3 & s_4 & s_5 & s_6 & s_7 & s_8 & s_9 & s_{10} \\
\hline
0 & 0 & 0 & 0 & 1 & 0 & 1 & 1 & 1 & 0 & 0 & -1 & -1 & -1 & 0 & -1 & 0 \\
0 & 0 & 1 & 1 & 2 & 0 & 0 & 1 & 1 & -1 & -1 & -1 & -1 & -1 & 0 & -1 & 0 \\
0 & 0 & 0 & 0 & 1 & 0 & 0 & 1 & 0 & -1 & 0 & 0 & -1 & -1 & 0 & -1 & 0 \\
0 & 0 & 0 & 0 & 1 & 0 & 0 & 0 & 1 & -1 & 0 & -1 & 0 & -1 & 0 & -1 & 0 \\
0 & 0 & 0 & 0 & 1 & 0 & 0 & 0 & 0 & -1 & 0 & 0 & 0 & -1 & 0 & -1 & 0 \\
0 & 0 & 0 & 0 & 1 & 0 & 0 & 0 & 1 & 0 & -1 & -1 & -1 & 0 & 0 & -1 & 0 \\
0 & 0 & 0 & 0 & 0 & 0 & 0 & 0 & 1 & 0 & 0 & -1 & 0 & 0 & 0 & -1 & 1 \\
0 & 0 & 0 & 0 & 1 & 0 & 0 & 1 & 0 & 0 & -1 & -1 & -1 & -1 & 0 & 0 & 0 \\
0 & 0 & 0 & 0 & 0 & 0 & 0 & 1 & 0 & 0 & 0 & 0 & -1 & -1 & 1 & 0 & 0 \\
0 & 0 & 0 & 0 & 1 & 0 & 0 & 0 & 0 & 0 & -1 & -1 & -1 & 0 & 0 & 0 & 0 \\
0 & 0 & 0 & 0 & 1 & 1 & 0 & 0 & 0 & -1 & -1 & 0 & 0 & 0 & 0 & 0 & 0 \\
1 & 1 & 0 & 0 & 0 & 0 & 0 & -1 & -1 & 0 & 0 & 0 & 0 & 0 & 0 & 0 & 0 
\end{array}
\right) 
$}
~,~
\eea
and
\beal{es30a06}
Q_{D} = 
\resizebox{0.58\textwidth}{!}{$
\left(
\begin{array}{ccccc|cc|cccccccccc}
 p_1 & p_2 & p_3 & p_4 & p_5 & q_1 & q_2 & s_1 & s_2 & s_3 & s_4 & s_5 & s_6 & s_7 & s_8 & s_9 & s_{10} \\
\hline
0 & 0 & 0 & 0 & 0 & 0 & 0 & 0 & 1 & 0 & 0 & 0 & 0 & 0 & 0 & -1 & 0 \\
0 & 0 & 0 & 0 & 0 & 0 & 0 & 0 & 0 & 0 & 1 & 0 & 0 & -1 & 0 & 0 & 0 \\
0 & 0 & 0 & 0 & 0 & 0 & 0 & 1 & 0 & 0 & 0 & 0 & -1 & 0 & 0 & 0 & 0 \\
0 & 0 & 0 & 0 & 0 & 0 & 0 & 0 & 0 & 1 & 0 & -1 & 0 & 0 & 0 & 0 & 0 \\
0 & 0 & 0 & 0 & 0 & 0 & 0 & 0 & 0 & 0 & -1 & 0 & 0 & 0 & 0 & 1 & 0 \\
0 & 0 & 0 & 0 & 0 & 0 & 0 & -1 & 0 & 0 & 0 & 0 & 0 & 1 & 0 & 0 & 0 \\
0 & 0 & 0 & 0 & 0 & 0 & 0 & 0 & 0 & -1 & 0 & 0 & 1 & 0 & 0 & 0 & 0 
\end{array}
\right) 
$}
~.~
\eea
The resulting toric diagram is summarized by the following $G_t$ matrix,
\beal{es30a06}
G_{t} = 
\resizebox{0.54\textwidth}{!}{$
\left(
\begin{array}{ccccc|cc|cccccccccc}
 p_1 & p_2 & p_3 & p_4 & p_5 & q_1 & q_2 & s_1 & s_2 & s_3 & s_4 & s_5 & s_6 & s_7 & s_8 & s_9 & s_{10} \\
\hline
1 & 1 & 1 & 1 & 1 & 1 & 1 & 1 & 1 & 1 & 1 & 1 & 1 & 1 & 1 & 1 & 1 \\
-1 & 1 & 0 & 0 & 0 & 0 & 0 & 0 & 0 & 0 & 0 & 0 & 0 & 0 & 0 & 0 & 0 \\
0 & 0 & -1 & 1 & 0 & 0 & 0 & 0 & 0 & 0 & 0 & 0 & 0 & 0 & 0 & 0 & 0 \\
0 & 0 & 0 & 2 & -1 & 1 & 1 & 0 & 0 & 0 & 0 & 0 & 0 & 0 & 0 & 0 & 0 \\
\end{array}
\right) 
$}
~,~
\eea
and is shown in \fref{f_toric_2a}.
Based on the $P$-matrix in \eref{es30a02}, global symmetry charge vectors $\alpha_\mu$ corresponding to the extremal brick matchings are assigned to the chiral and Fermi fields of the theory as shown in \tref{t_global_2a}.
The sum of the charge vectors $\alpha_\mu$ vanishes,
\beal{es30a06b}
\alpha_1 + \alpha_2 + \alpha_3 + \alpha_4 + \alpha_5 = 0 ~,~
\eea
in order for the $J$- and $E$-terms to be invariant under the global symmetry.

\begin{table}[H]
\begin{center}
\begin{tabular}{|p{0.6cm}|p{1.2cm}|}
\hline
$P_{51}$ & $\alpha_3$\\
$P_{62}$ & $\alpha_3$\\
$P_{73}$ & $\alpha_3$\\
$P_{84}$ & $\alpha_3$\\
$Q_{51}$ & $\alpha_4$\\
$Q_{62}$ & $\alpha_4$\\
$Q_{73}$ & $\alpha_4$\\
$Q_{84}$ & $\alpha_4$\\
\hline
\end{tabular}
~
\begin{tabular}{|p{0.6cm}|p{1.2cm}|}
\hline
$U_{25}$ & $\alpha_4$\\
$U_{47}$ & $\alpha_4$\\
$W_{25}$ & $\alpha_3$\\
$W_{47}$ & $\alpha_3$\\
$X_{14}$ & $\alpha_2$\\
$X_{32}$ & $\alpha_2$\\
$X_{58}$ & $\alpha_2$\\
$X_{76}$ & $\alpha_2$\\
\hline
\end{tabular}
~
\begin{tabular}{|p{0.6cm}|p{1.2cm}|}
\hline
$Y_{14}$ & $\alpha_1$\\
$Y_{21}$ & $\alpha_5$\\
$Y_{32}$ & $\alpha_1$\\
$Y_{43}$ & $\alpha_5$\\
$Y_{58}$ & $\alpha_1$\\
$Y_{65}$ & $\alpha_5$\\
$Y_{76}$ & $\alpha_1$\\
$Y_{87}$ & $\alpha_5$\\
\hline
\end{tabular}
~
\begin{tabular}{|p{0.6cm}|p{1.2cm}|}
\hline
$\Lambda_{21}$ & $\alpha_3 +\alpha_4$\\
$\Lambda_{43}$ & $\alpha_3 +\alpha_4$\\
$\Lambda_{65}$ & $\alpha_3 +\alpha_4$\\
$\Lambda_{87}$ & $\alpha_3 +\alpha_4$\\
$\Lambda_{54}^1$ & $\alpha_1 +\alpha_4$\\
$\Lambda_{54}^2$ & $\alpha_2 +\alpha_4$\\
$\Lambda_{54}^3$ & $\alpha_2 +\alpha_3$\\
$\Lambda_{54}^4$ & $\alpha_1 +\alpha_3$\\
\hline
\end{tabular}
~
\begin{tabular}{|p{0.6cm}|p{1.2cm}|}
\hline
$\Lambda_{72}^1$ & $\alpha_2 +\alpha_4$\\
$\Lambda_{72}^2$ & $\alpha_2 +\alpha_3$\\
$\Lambda_{72}^3$ & $\alpha_1 +\alpha_3$\\
$\Lambda_{72}^4$ & $\alpha_1 +\alpha_4$\\
$\Lambda_{61}^1$ & $\alpha_4 +\alpha_5$\\
$\Lambda_{61}^2$ & $\alpha_3 +\alpha_5$\\
$\Lambda_{83}^1$ & $\alpha_4 +\alpha_5$\\
$\Lambda_{83}^2$ & $\alpha_3 +\alpha_5$\\
\hline
\end{tabular}
\caption{
Global symmetry charge assignments for chiral and Fermi fields of the $\mathbb{F}_{0,+-}$ theory.
\label{t_global_2a}}
 \end{center}
 \end{table}

Now, let us introduce the following mass deformation,
\beal{es30a10}
\begin{array}{rrrclcrcl}
& & & J  & & &  E & \textcolor{blue}{+}  & \textcolor{blue}{\Delta E} \\
\Lambda_{21} : & \ \ \ & Y_{14} \cdot Y_{43} \cdot X_{32}  &-&  X_{14} \cdot Y_{43} \cdot Y_{32}   & \ \ \ \ &  \textcolor{blue}{-Y_{21}} + U_{25} \cdot P_{51}  &-&  W_{25} \cdot Q_{51}  \\
\Lambda_{43} : & \ \ \ & X_{32} \cdot Y_{21} \cdot Y_{14}  &-&  Y_{32} \cdot Y_{21} \cdot X_{14}   & \ \ \ \ &  \textcolor{blue}{+Y_{43}} + W_{47} \cdot  Q_{73}  &-&  U_{47} \cdot P_{73}   \\
\Lambda_{65} : & \ \ \ & Y_{58} \cdot Y_{87} \cdot X_{76}  &-&  X_{58} \cdot Y_{87} \cdot Y_{76}   & \ \ \ \ &  \textcolor{blue}{+Y_{65}} + Q_{62} \cdot  W_{25}  &-&  P_{62} \cdot U_{25}  \\
 \Lambda_{87} : & \ \ \ & X_{76} \cdot Y_{65} \cdot Y_{58}  &-&  Y_{76} \cdot Y_{65} \cdot X_{58}   & \ \ \ \ &  \textcolor{blue}{-Y_{87}} + P_{84} \cdot U_{47}  &-&  Q_{84} \cdot  W_{47}   \\
\end{array}
~.~
\eea
Integrating out the massive fields, the massive chiral fields are replaced as follows
\beal{es30a11}
Y_{21} & = U_{25} \cdot P_{51} - W_{25} \cdot Q_{51} ~,~ 
Y_{43} & = U_{47} \cdot P_{73} - W_{47} \cdot  Q_{73} ~,~  \nn\\
Y_{65} & = P_{62} \cdot U_{25} - Q_{62} \cdot  W_{25} ~,~ 
Y_{87} & = P_{84} \cdot U_{47} - Q_{84} \cdot  W_{47} ~.~
\eea
We note that the massive chiral fields $Y_{21}$, $Y_{43}$, $Y_{65}$, $Y_{87}$ and the massive Fermi fields $\bar{\Lambda}_{21}$, $\bar{\Lambda}_{43}$, $\bar{\Lambda}_{65}$, $\bar{\Lambda}_{87}$ are all part of the brick matching $p_5$ highlighted in blue in the $P$-matrix in \eref{es30a02}.

\begin{figure}[H]
\begin{center}
\resizebox{0.35\hsize}{!}{
\includegraphics[height=6cm]{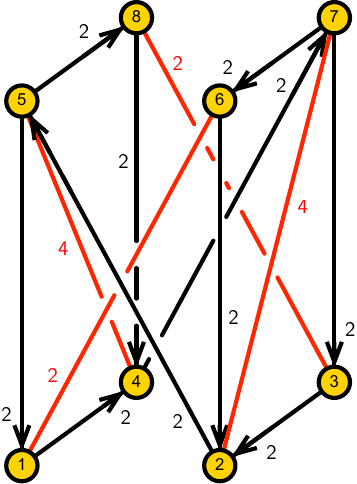} 
}
\caption{
The quiver obtained from the $\mathbb{F}_{0,+-}$ theory after the mass deformation in \eqref{es30a10}.
\label{f_quiver_2b}}
 \end{center}
 \end{figure}

The quiver diagram obtained after integrating the massive fields is shown in \fref{f_quiver_2b}. The $J$- and $E$-terms become
\beal{es30a20}
\resizebox{0.95\textwidth}{!}{$
\begin{array}{rrrclcrcl}
& & & J  & & &  & E  \\
 \Lambda^1_{54} : & \ \ \ & W_{47} \cdot X_{76} \cdot P_{62} \cdot U_{25}  &-&  U_{47} \cdot X_{76} \cdot P_{62} \cdot W_{25}   & \ \ \ \ &  Q_{51} \cdot Y_{14}  &-&  Y_{58} \cdot  Q_{84}   \\
 \Lambda^2_{54} : & \ \ \ & W_{47} \cdot Y_{76} \cdot P_{62} \cdot U_{25}  &-&  U_{47} \cdot Y_{76} \cdot P_{62} \cdot W_{25}   & \ \ \ \ &  Q_{51} \cdot X_{14}  &-&  X_{58} \cdot  Q_{84}   \\
 \Lambda^3_{54} : & \ \ \ & W_{47} \cdot Y_{76} \cdot Q_{62} \cdot U_{25}  &-&  U_{47} \cdot Y_{76} \cdot Q_{62} \cdot W_{25}   & \ \ \ \ &  P_{51} \cdot X_{14}  &-&  X_{58} \cdot  P_{84}   \\
 \Lambda^4_{54} : & \ \ \ & W_{47} \cdot X_{76} \cdot Q_{62} \cdot U_{25}  &-&  U_{47} \cdot X_{76} \cdot Q_{62} \cdot W_{25}   & \ \ \ \ &  P_{51} \cdot Y_{14}  &-&  Y_{58} \cdot  P_{84}   \\
 \Lambda^1_{72} : & \ \ \ & W_{25} \cdot Y_{58} \cdot P_{84} \cdot U_{47}  &-&  U_{25} \cdot Y_{58} \cdot P_{84} \cdot W_{47}   & \ \ \ \ &  Q_{73} \cdot X_{32}  &-&  X_{76} \cdot  Q_{62}   \\
 \Lambda^2_{72} : & \ \ \ & W_{25} \cdot Y_{58} \cdot Q_{84} \cdot U_{47}  &-&  U_{25} \cdot Y_{58} \cdot Q_{84} \cdot W_{47}   & \ \ \ \ &  P_{73} \cdot X_{32}  &-&  X_{76} \cdot  P_{62}   \\
 \Lambda^3_{72} : & \ \ \ &  W_{25} \cdot X_{58} \cdot Q_{84} \cdot U_{47}  &-&  U_{25} \cdot X_{58} \cdot Q_{84} \cdot W_{47}   & \ \ \ \ &  P_{73} \cdot Y_{32}  &-&  Y_{76} \cdot  P_{62}   \\
 \Lambda^4_{72} : & \ \ \ & W_{25} \cdot X_{58} \cdot P_{84} \cdot U_{47}  &-&  U_{25} \cdot X_{58} \cdot P_{84} \cdot W_{47}   & \ \ \ \ &  Q_{73} \cdot Y_{32}  &-&  Y_{76} \cdot  Q_{62}   \\
 \Lambda^1_{61} : & \ \ \ &  Y_{14} \cdot W_{47} \cdot X_{76} &-&   X_{14} \cdot W_{47} \cdot Y_{76}   & \ \ \ \ &  Q_{62} \cdot U_{25} \cdot P_{51}  &-&  P_{62} \cdot U_{25} \cdot  Q_{51}   \\
 \Lambda^2_{61} : & \ \ \ & Y_{14} \cdot U_{47} \cdot X_{76} &-&   X_{14} \cdot U_{47} \cdot Y_{76}   & \ \ \ \ &  Q_{62} \cdot W_{25} \cdot P_{51}  &-&  P_{62} \cdot W_{25} \cdot  Q_{51}   \\
\Lambda^1_{83} : & \ \ \ &  Y_{32} \cdot W_{25} \cdot X_{58} &-&   X_{32} \cdot W_{25} \cdot Y_{58}   & \ \ \ \ &  Q_{84} \cdot U_{47} \cdot P_{73}  &-&  P_{84} \cdot U_{47} \cdot  Q_{73}   \\
 \Lambda^2_{83} : & \ \ \ &  Y_{32} \cdot U_{25} \cdot X_{58} &-&   X_{32} \cdot U_{25} \cdot Y_{58}   & \ \ \ \ &  Q_{84} \cdot W_{47} \cdot P_{73}  &-&  P_{84} \cdot W_{47} \cdot  Q_{73}   \\
\end{array}
$}
~.~
\nn\\
\eea
This theory is one of the toric phases of $Q^{1,1,1}/\mathbb{Z}_2$ \cite{Franco:2018qsc}. Anyhow, let us study it in detail.

\begin{figure}[H]
\begin{center}
\resizebox{0.4\hsize}{!}{
\includegraphics[height=6cm]{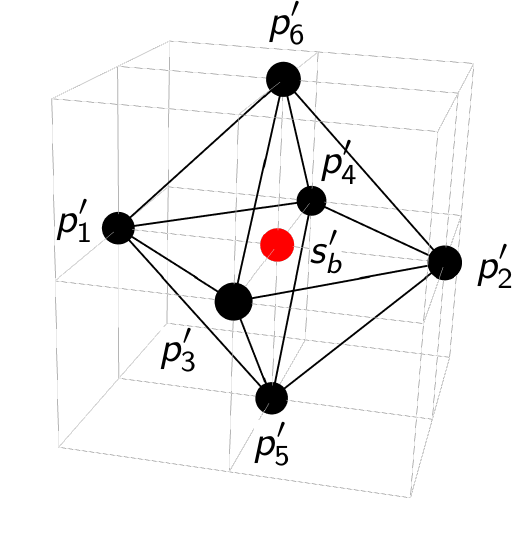} 
}
\caption{The toric diagram of $Q^{1,1,1}/\mathbb{Z}_2$ obtained from a mass deformation of the $\mathbb{F}_{0,+-}$ theory.
\label{f_toric_2b}}
 \end{center}
 \end{figure}

The corresponding $P$-matrix can be obtained via the forward algorithm as follows,
\beal{es30a21}
P_{\Lambda} = 
\resizebox{0.5\textwidth}{!}{$
\left(
\begin{array}{c|cccccc|cccccccccc}
&p'_1 & p'_2 & p'_3 & p'_4 & p'_5 & p'_6 & s'_1 & s'_2 & s'_3 & s'_4 & s'_5 & s'_6 & s'_7 & s'_8 & s'_9 & s'_{10} \\
\hline
P_{51} & 0 & 0 & 0 & 1 & 0 & 0 & 0 & 0 & 0 & 0 & 0 & 0 & 0 & 0 & 1 & 1 \\ 
P_{62} & 0 & 0 & 0 & 1 & 0 & 0 & 0 & 0 & 0 & 0 & 0 & 0 & 1 & 1 & 0 & 0 \\ 
P_{73} & 0 & 0 & 0 & 1 & 0 & 0 & 0 & 0 & 0 & 0 & 0 & 1 & 0 & 1 & 0 & 0 \\
P_{84} & 0 & 0 & 0 & 1 & 0 & 0 & 0 & 0 & 0 & 0 & 1 & 0 & 0 & 0 & 0 & 1 \\
Q_{51} & 0 & 0 & 1 & 0 & 0 & 0 & 0 & 0 & 0 & 0 & 0 & 0 & 0 & 0 & 1 & 1 \\ 
Q_{62} & 0 & 0 & 1 & 0 & 0 & 0 & 0 & 0 & 0 & 0 & 0 & 0 & 1 & 1 & 0 & 0 \\
Q_{73} & 0 & 0 & 1 & 0 & 0 & 0 & 0 & 0 & 0 & 0 & 0 & 1 & 0 & 1 & 0 & 0 \\ 
Q_{84} & 0 & 0 & 1 & 0 & 0 & 0 & 0 & 0 & 0 & 0 & 1 & 0 & 0 & 0 & 0 & 1 \\
U_{25} & 0 & 1 & 0 &  0 & 0 & 0 & 0 & 0 & 0 & 1 & 0 & 0 & 0 & 0 & 0 & 0 \\
U_{47} & 0 & 1 & 0 & 0 & 0 & 0 & 0 & 0 & 1 & 0 & 0 & 0 & 0 & 0 & 0 & 0 \\ 
W_{25} & 1 & 0 & 0 & 0 & 0 & 0 & 0 & 0 & 0 & 1 & 0 & 0 & 0 & 0 & 0 & 0 \\
W_{47} & 1 & 0 & 0 & 0 & 0 & 0 & 0 & 0 & 1 & 0 & 0 & 0 & 0 & 0 & 0 & 0 \\ 
X_{14} & 0 & 0 & 0 & 0 & 1 & 0 & 0 & 1 & 0 & 0 & 1 & 0 & 0 & 0 & 0 & 0 \\ 
X_{32} & 0 & 0 & 0 & 0 & 1 & 0 & 1 & 0 & 0 & 0 & 0 & 0 & 1 & 0 & 0 & 0 \\ 
X_{58} & 0 & 0 & 0 & 0 & 1 & 0 & 0 & 1 & 0 & 0 & 0 & 0 & 0 & 0 & 1 & 0 \\ 
X_{76} & 0 & 0 & 0 & 0 & 1 & 0 & 1 & 0 & 0 & 0 & 0 & 1 & 0 & 0 & 0 & 0 \\ 
Y_{14} & 0 & 0 & 0 & 0 & 0 & 1 & 0 & 1 & 0 & 0 & 1 & 0 & 0 & 0 & 0 & 0 \\ 
Y_{32} & 0 & 0 & 0 & 0 & 0 & 1 & 1 & 0 & 0 & 0 & 0 & 0 & 1 & 0 & 0 & 0 \\ 
Y_{58} & 0 & 0 & 0 & 0 & 0 & 1 & 0 & 1 & 0 & 0 & 0 & 0 & 0 & 0 & 1 & 0 \\
Y_{76} & 0 & 0 & 0 & 0 & 0 & 1 & 1 & 0 & 0 & 0 & 0 & 1 & 0 & 0 & 0 & 0 \\ 
\hline
\Lambda_{54} & 0 & 0 & 1 & 0 & 0 & 1 & 0 & 1 & 0 & 0 & 1 & 0 & 0 & 0 & 1 & 1 \\ 
\Lambda^2_{54} & 0 & 0 & 1 & 0 & 1 & 0 & 0 & 1 & 0 & 0 & 1 & 0 & 0 & 0 & 1 & 1 \\
\Lambda^3_{54} & 0 & 0 & 0 & 1 & 1 & 0 & 0 & 1 & 0 & 0 & 1 & 0 & 0 & 0 & 1 & 1 \\
\Lambda^4_{54} & 0 & 0 & 0 & 1 & 0 & 1 & 0 & 1 & 0 & 0 & 1 & 0 & 0 & 0 & 1 & 1 \\
\Lambda_{72} & 0 & 0 & 1 & 0 & 1 & 0 & 1 & 0 & 0 & 0 & 0 & 1 & 1 & 1 & 0 &  0 \\
\Lambda^2_{72} & 0 & 0 & 0 & 1 & 1 & 0 & 1 & 0 & 0 & 0 & 0 & 1 & 1 & 1 & 0 & 0 \\
\Lambda^3_{72} & 0 & 0 & 0 & 1 & 0 & 1 & 1 & 0 & 0 & 0 & 0 & 1 & 1 & 1 & 0 & 0 \\ 
\Lambda^4_{72} & 0 & 0 & 1 & 0 & 0 & 1 & 1 & 0 & 0 & 0 & 0 & 1 & 1 & 1 & 0 & 0 \\ 
\Lambda_{61} & 0 & 1 & 1 & 1 & 0 & 0 & 0 & 0 & 0 & 1 & 0 & 0 & 1 & 1 & 1 & 1 \\ 
\Lambda^2_{61} & 1 & 0 & 1 & 1 & 0 & 0 & 0 & 0 & 0 & 1 & 0 & 0 & 1 & 1 & 1 & 1 \\
\Lambda_{83} & 0 & 1 & 1 & 1 & 0 & 0 & 0 & 0 & 1 & 0 & 1 & 1 & 0 & 1 & 0 & 1 \\ 
\Lambda^2_{83} & 1 & 0 & 1 & 1 & 0 & 0 & 0 & 0 & 1 & 0 & 1 & 1 & 0 & 1 & 0 & 1
\end{array}
\right) 
$} 
~,~
\eea
where we have called the new brick matchings $p'_\mu$ to distinguish them from the original ones.
The $Q_{JE}$ and $Q_D$ matrices are, respectively, 
\beal{es30a22}
Q_{JE} = 
\resizebox{0.53\textwidth}{!}{$
\left(
\begin{array}{cccccc|cccccccccc}
p'_1 & p'_2 & p'_3 & p'_4 & p'_5 & p'_6 & s'_1 & s'_2 & s'_3 & s'_4 & s'_5 & s'_6 & s'_7 & s'_8 & s'_9 & s'_{10} \\
\hline
0 & 0 & 1 & 1 & 0 & 0 & 1 & 1 & 0 & 0 & -1 & -1 & -1 & 0 & -1 & 0 \\ 
0 & 0 & 0 & 0 & 0 & 0 & 0 & 1 & 0 & 0 & -1 & 0 & 0 & 0 & -1 & 1 \\ 
0 & 0 & 0 & 0 & 0 & 0 & 1 & 0 & 0 & 0 & 0 & -1 & -1 & 1 & 0 & 0 \\ 
1 & 1 & 0 & 0 & 0 & 0 & 0 & 0 & -1 & -1 & 0 & 0 & 0 & 0 & 0 & 0 \\ 
0 & 0 & 0 & 0 & 1 & 1 & -1 & -1 & 0 & 0 & 0 & 0 & 0 & 0 & 0 & 0
\end{array}
\right) 
$}
~,~
\eea
and
\beal{es30a22b}
Q_{D} = 
\resizebox{0.52\textwidth}{!}{$
\left(
\begin{array}{cccccc|cccccccccc}
p'_1 & p'_2 & p'_3 & p'_4 & p'_5 & p'_6 & s'_1 & s'_2 & s'_3 & s'_4 & s'_5 & s'_6 & s'_7 & s'_8 & s'_9 & s'_{10} \\
\hline
0 & 0 & 0 & 0 & 0 & 0 & 0 & 1 & 0 & 0 & 0 & 0 & 0 & 0 & -1 & 0 \\ 
0 & 0 & 0 & 0 & 0 & 0 & 0 & 0 & 0 & 1 & 0 & 0 & -1 & 0 & 0 & 0 \\ 
0 & 0 & 0 & 0 & 0 & 0 & 1 & 0 & 0 & 0 &  0 & -1 & 0 & 0 & 0 & 0 \\ 
0 & 0 & 0 & 0 & 0 & 0 & 0 & 0 & 1 & 0 & -1 & 0 & 0 & 0 & 0 &  0 \\ 
0 & 0 & 0 & 0 & 0 & 0 & 0 & 0 & 0 & -1 & 0 & 0 & 0 & 0 & 1 & 0 \\ 
0 & 0 & 0 & 0 &  0 & 0 & -1 & 0 & 0 & 0 & 0 & 0 & 1 & 0 & 0 & 0 \\ 
0 & 0 & 0 & 0 & 0 & 0 & 0 & 0 & -1 & 0 & 0 & 1 & 0 & 0 & 0 & 0
\end{array}
\right) 
$}
~.~
\eea
The resulting toric is encoded in the following matrix and 
\beal{es30a22b}
G_{t} = 
\resizebox{0.49\textwidth}{!}{$
\left(
\begin{array}{cccccc|cccccccccc}
p'_1 & p'_2 & p'_3 & p'_4 & p'_5 & p'_6 & s'_1 & s'_2 & s'_3 & s'_4 & s'_5 & s'_6 & s'_7 & s'_8 & s'_9 & s'_{10} \\
\hline
 1 & 1 & 1 & 1 & 1 & 1 & 1 & 1 & 1 & 1 & 1 & 1 & 1 & 1 & 1 & 1 \\ 
 -1 & 1 & 0 & 0 & 0 & 0 & 0 & 0 & 0 & 0 & 0 & 0 & 0 & 0 & 0 & 0 \\ 
 0 & 0 & -1 & 1 & 0 & 0 & 0 & 0 & 0 & 0 & 0 & 0 & 0 & 0 & 0 & 0\\
  0 & 0 & 0 & 0 & -1 & 1 & 0 & 0 & 0 & 0 & 0 & 0 & 0 & 0 & 0 & 0 \\ 
\end{array}
\right) 
$}
~,~
\eea
and shown in \fref{f_toric_2b}. This is indeed the toric diagram for $Q^{1,1,1}/\mathbb{Z}_2$, as expected.

\begin{figure}[H]
\begin{center}
\resizebox{0.8\hsize}{!}{
\includegraphics[height=6cm]{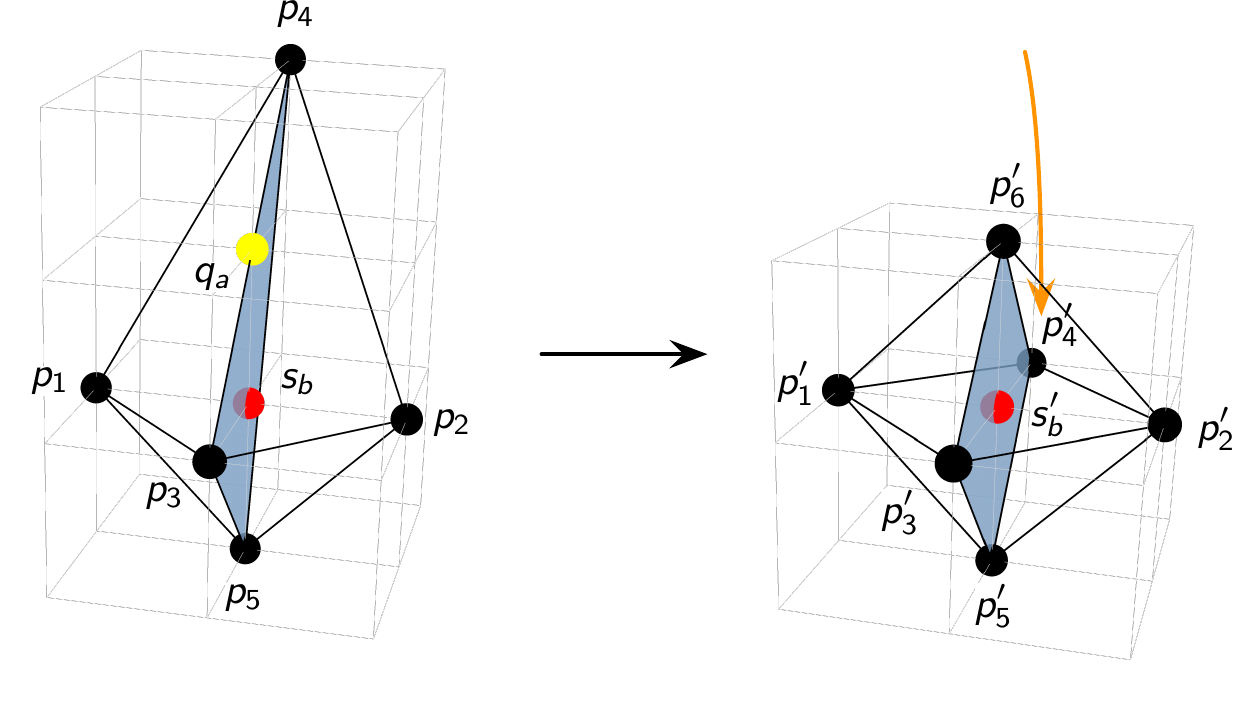} 
}
\caption{
The toric diagram of $\mathbb{F}_{0,+-}$ changing into the toric diagram of $Q^{1,1,1}/\mathbb{Z}_2$ under mass deformation. We note that brick matchings $p_3$, $p_4$ and $p_5$ define a 2-dimensional plane (blue) and correspond to the massive terms added to the $J$- and $E$-terms. On this plane, the point associated to $p_4$ moves to a new position due to the mass deformation.
\label{f_toric_2a3}}
 \end{center}
 \end{figure}

A comparison between the toric diagrams before and after the mass deformation reveals that most of the points in them preserve their positions. Only the point corresponding to the extremal brick matching $p_4$ in $\mathbb{F}_{0,+-}$, which is highlighted in orange in \eref{es30a02}, moves. In this process, $q_1$ and $q_2$ become a single extremal brick matching denoted as $p'_6$. This is illustrated in \fref{f_toric_2a3}.

As shown in \fref{f_toric_2a3}, the massive brick matching $p_5$, the unaffected brick matching $p_3$ and the moving brick matching $p_4$ are on the mass deformation plane that dissects the toric diagram. 
Brick matchings $p_5 \mapsto p'_5$, $p_3 \mapsto p'_3$, $p_4 \mapsto p'_4$ and the new extremal brick matching $q_1,q_2 \mapsto p'_6$ are all on the same plane that stays invariant during the mass deformation.
The mass deformation plane is highlighted in blue in \fref{f_toric_2a3}.
Under the deformation, the point in the toric diagram corresponding to the brick matching $p_4$ moves along this plane to its new position $p'_4$.

According to the global symmetry charges for individual fields in \tref{t_global_2a},
the mass terms in \eref{es30a10},
\beal{es30a25}
\bar{\Lambda}_{21} \cdot Y_{21} ~,~
\bar{\Lambda}_{43} \cdot Y_{43} ~,~
\bar{\Lambda}_{65} \cdot Y_{65} ~,~
\bar{\Lambda}_{87} \cdot Y_{87} ~,~
\eea
carry the following charges,
\beal{es30a26}
&
\alpha(\bar{\Lambda}_{21} \cdot Y_{21}) = \alpha_5 - \alpha_3 - \alpha_4 ~,~
\alpha(\bar{\Lambda}_{43} \cdot Y_{43}) = \alpha_5 - \alpha_3 - \alpha_4 ~,~
&
\nn\\
&
\alpha(\bar{\Lambda}_{65} \cdot Y_{65}) = \alpha_5 - \alpha_3 - \alpha_4~,~
\alpha(\bar{\Lambda}_{87} \cdot Y_{87}) = \alpha_5 - \alpha_3 - \alpha_4~.~
&
\eea
We see that the four mass terms carry the same non-zero global symmetry charge. These charges correspond to the extremal brick matchings $p_3$, $p_4$, and $p_5$ whose associated points in the toric diagram are exactly those on the mass deformation plane shown in \fref{f_toric_2a3}. The mass terms break part of the global symmetry of the original $\mathbb{F}_{0,+-}$ theory. Geometrically, this symmetry breaking occurs along the mass deformation plane. The global symmetry of $Q^{1,1,1}/\mathbb{Z}_2$ contains linear combinations of the original symmetries which are orthogonal to $\alpha_5 - \alpha_3 - \alpha_4$.

As for the previous example, let us compare the minimum volumes of the underlying Sasaki-Einstein 7-manifolds. 
They are 
\beal{es30a30}
V_{min}^{\mathbb{F}_{0,+-}} = \frac{45+26\sqrt{3}}{576} ~,~
V_{min}^{Q^{1,1,1}/ \mathbb{Z}_2} \simeq 0.150916
\eea
The ratio between the two volumes is 
\beal{es30a31} 
\frac{V_{min}^{\mathbb{F}_{0,+-}}}{V_{min}^{Q^{1,1,1}/ \mathbb{Z}_2} } 
\simeq
0.870919 < 1
~,~
\eea
which shows that the minimum volume increases with mass deformation, as expected. 

The mass deformation connecting $\mathbb{F}_{0,+-}$ to $Q^{1,1,1}/ \mathbb{Z}_2$ that we discussed in this section is related, via orbifold reduction, to the one connecting the $4d$ theories for $\mathbb{C}^3/\mathbb{Z}_4$ and $\mathbb{F}_0$. The toric diagrams for these two CY$_3$'s arise on the mass deformation plane, as shown in \fref{f_toric_2a3}.
\\

\subsection{Mass Deformation of the $\mathbb{C}^{2}/\mathbb{Z}_{3} \times \mathbb{C}^2 $ Theory \label{sec:43}}

\begin{figure}[H]
\begin{center}
\resizebox{0.35\hsize}{!}{
\includegraphics[height=6cm]{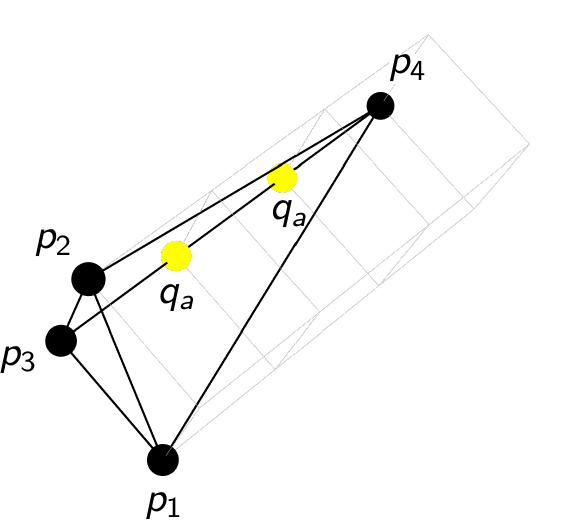} 
}
\caption{
The toric diagram for $\mathbb{C}^{2}/\mathbb{Z}_{3} \times \mathbb{C}^2$.
\label{f_toric_3a}}
 \end{center}
 \end{figure}

Finally, let us consider $\mathbb{C}^{2}/\mathbb{Z}_{3} \times \mathbb{C}^2 $ whose toric diagram is shown in \fref{f_toric_3a}.

\begin{figure}[H]
\begin{center}
\resizebox{0.35\hsize}{!}{
\includegraphics[height=6cm]{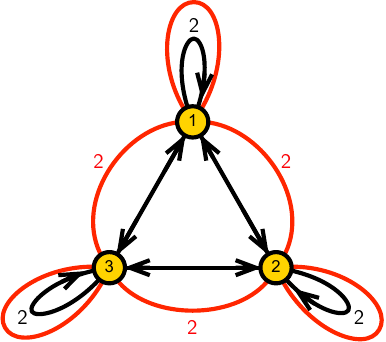} 
}
\caption{
The quiver of the $\mathbb{C}^{2}/\mathbb{Z}_{3} \times \mathbb{C}^2$ theory.
\label{f_quiver_3a}}
 \end{center}
 \end{figure}

The corresponding $2d$ gauge theory can be constructed via standard orbifold techniques and has $(4,4)$ SUSY. Its quiver is shown in \fref{f_quiver_3a}.

The $J$- and $E$-terms are
\beal{es40a01}
\begin{array}{rrrclcrcl}
& & & J  & & & & E   \\
\Lambda_{12 } : & \ \ \ & X_{22} \cdot X_{21}  &-&  X_{21} \cdot X_{11}  & \ \ \ \ &Z_{11} \cdot X_{12}  &-&  X_{12} \cdot Z_{22}  \\
\Lambda_{21} : & \ \ \ & X_{12} \cdot X_{22}  &-&  X_{11} \cdot X_{12}  & \ \ \ \ & Z_{22} \cdot X_{21}  &-&  X_{21} \cdot Z_{11} \\
\Lambda_{13} : & \ \ \ &  X_{31} \cdot X_{11}  &-&  X_{33} \cdot X_{31}  & \ \ \ \ & Z_{11} \cdot X_{13}  &-&  X_{13} \cdot  Z_{33} \\
\Lambda_{31} : & \ \ \ & X_{11} \cdot X_{13}   &-& X_{13} \cdot X_{33}  & \ \ \ \ & Z_{33} \cdot X_{31}  &-&  X_{31} \cdot Z_{11}  \\
\Lambda_{23} : & \ \ \ & X_{33} \cdot X_{32}  &-&  X_{32} \cdot X_{22}  & \ \ \ \ & Z_{22} \cdot X_{23}  &-&  X_{23} \cdot Z_{33} \\
\Lambda_{32} : & \ \ \ & X_{23} \cdot X_{33}  &-&  X_{22} \cdot X_{23}  & \ \ \ \ &  Z_{33} \cdot X_{32}  &-&  X_{32} \cdot Z_{22} \\
\Lambda_{11} : & \ \ \ & X_{13} \cdot X_{31}  &-&  X_{12} \cdot X_{21}  & \ \ \ \ & Z_{11} \cdot X_{11}  &-&  X_{11} \cdot  Z_{11} \\
\Lambda_{22} : & \ \ \ & X_{21} \cdot X_{12}  &-&  X_{23} \cdot X_{32}  & \ \ \ \ & Z_{22} \cdot X_{22}  &-&  X_{22} \cdot  Z_{22} \\
\Lambda_{33} : & \ \ \ & X_{32} \cdot X_{23}  &-&  X_{31} \cdot X_{13}  & \ \ \ \ & Z_{33} \cdot X_{33}  &-&  X_{33} \cdot  Z_{33} 
\end{array}
 ~,~
\eea

The $P$-matrix becomes
\beal{es40a02}
P_\Lambda=
\resizebox{0.38\textwidth}{!}{$
\left(
\begin{array}{c|>{\columncolor{Blue!20}}ccc>{\columncolor{Orange!20}}c|cccccc}
\; & p_1 & p_2 & p_3 & p_4 & q_1 & q_2 & q_3 & q_4 & q_5 & q_6  \\
\hline
X_{11} & 1 & 0 & 0 & 0 & 0 & 0 & 0 & 0 & 0 & 0 \\ 
X_{22} & 1 & 0 & 0 & 0 & 0 & 0 & 0 & 0 & 0 & 0 \\ 
X_{33} & 1 & 0 & 0 & 0 & 0 & 0 & 0 & 0 & 0 & 0 \\ 
X_{12} & 0 & 0 & 1 & 0 & 0 & 1 & 1 & 0 & 0 & 1 \\ 
X_{13} & 0 & 0 & 0 & 1 & 0 & 0 & 1 & 0 & 1 & 1 \\ 
X_{21} & 0 & 0 & 0 & 1 & 1 & 0 & 0 & 1 & 1 & 0 \\ 
X_{23} & 0 & 0 & 1 & 0 & 1 & 0 & 1 & 0 & 1 & 0 \\ 
X_{31} & 0 & 0 & 1 & 0 & 1 & 1 & 0 & 1 & 0 & 0 \\ 
X_{32} & 0 & 0 & 0 & 1 & 0 & 1 & 0 & 1 & 0 & 1 \\ 
Z_{11} & 0 & 1 & 0 & 0 & 0 & 0 & 0 & 0 & 0 & 0 \\ 
Z_{22} & 0 & 1 & 0 & 0 & 0 & 0 & 0 & 0 & 0 & 0 \\ 
Z_{33} & 0 & 1 & 0 & 0 & 0 & 0 & 0 & 0 & 0 & 0 \\ 
\hline
\Lambda_{12} & 0 & 1 & 1 & 0 & 0 & 1 & 1 & 0 & 0 & 1 \\ 
\Lambda_{13} & 0 & 1 & 0 & 1 & 0 & 0 & 1 & 0 & 1 & 1 \\ 
\Lambda_{21} & 0 & 1 & 0 & 1 & 1 & 0 & 0 & 1 & 1 & 0 \\ 
\Lambda_{23} & 0 & 1 & 1 & 0 & 1 & 0 & 1 & 0 & 1 & 0 \\ 
\Lambda_{31} & 0 & 1 & 1 & 0 & 1 & 1 & 0 & 1 & 0 & 0 \\
\Lambda_{32} & 0 & 1 & 0 & 1 & 0 & 1 & 0 & 1 & 0 & 1 \\
\Lambda_{11} & 1 & 1 & 0 & 0 & 0 & 0 & 0 & 0 & 0 & 0 \\ 
\Lambda_{22} & 1 & 1 & 0 & 0 & 0 & 0 & 0 & 0 & 0 & 0 \\ 
 \Lambda_{33} & 1 & 1 & 0 & 0 & 0 & 0 & 0 & 0 & 0 & 0
\end{array}
\right)
$}
~.~
\eea

The charge matrices of GLSM fields that encode the $J$-, $E$- and $D$-terms are 
\beal{es40a05}
Q_{JE} = 
\resizebox{0.35\textwidth}{!}{$
\left(
\begin{array}{cccc|cccccc}
p_1 & p_2 & p_3 & p_4 & q_1 & q_2 & q_3 & q_4 & q_5 & q_6 \\
\hline
0 & 0 & 0 & -1 & -1 & -1 & 1 & 2 & 0 & 0 \\
0 & 0 & 0 & -1 & 0 & -1 & 0 & 1 & 0 & 1 \\ 
0 & 0 & 1 & 0 & -1 & -1 & 0 & 1 & 0 & 0 \\ 
0 & 0 & 0 & -1 & -1 & 0 & 0 & 1 & 1 & 0
 \end{array}
\right)
$}
~,~
Q_{D} = 
\resizebox{0.34\textwidth}{!}{$
\left(
\begin{array}{cccc|cccccc}
p_1 & p_2 & p_3 & p_4 & q_1 & q_2 & q_3 & q_4 & q_5 & q_6 \\
\hline
0 & 0 & 0 & 1 & 0 & 1 & 0 & -2 & 0 & 0 \\
0 & 0 & 0 & 0 & 1 & -1 & 0 & 0 & 0 & 0
 \end{array}
\right) 
$}
~.~
\nn\\
\eea

The resulting toric diagram is is given by the following matrix
\beal{es40a07}
G_{t} = 
\left(
\begin{array}{cccc|cccccc}
p_1 & p_2 & p_3 & p_4 & q_1 & q_2 & q_3 & q_4 & q_5 & q_6 \\
\hline
1 & 1 & 1 & 1 & 1 & 1 & 1 & 1 & 1 & 1 \\ 
1 & 0 & 0 & 0 & 0 & 0 & 0 & 0 & 0 & 0 \\ 
0 & 0 & 0 & 3 & 1 & 1 & 1 & 2 & 2 & 2 \\ 
0 & 1 & 0 & 0 & 0 & 0 & 0 & 0 & 0 & 0
\end{array}
\right) 
~.~
\eea
It is shown \fref{f_toric_3a} and it corresponds to $\mathbb{C}^{2}/\mathbb{Z}_{3} \times \mathbb{C}^2$, as expected.

Using the $P$-matrix in \eref{es40a02}, we can assign global symmetry charges $\alpha_\mu$ corresponding to extremal brick matchings $p_\mu$ to the chiral and Fermi fields of the brane brick model, as summarized in \tref{t_global_3a}.
Since every $J$- and $E$-term contains every extremal brick matching, invariance under the global symmetry requires that
\beal{es40a06b}
\alpha_1 + \alpha_2 + \alpha_3 + \alpha_4 = 0 ~.~
\eea

\begin{table}[H]
\begin{center}
\begin{tabular}{|c|p{1.3cm}|}
\hline
$X_{11}$ & $\alpha_1$\\
$X_{22}$ & $\alpha_1$\\
$X_{33}$ & $\alpha_1$\\
$X_{12}$ & $\alpha_3$\\
$X_{13}$ & $\alpha_4$\\
$X_{21}$ & $\alpha_4$\\
\hline
\end{tabular}
~
\begin{tabular}{|c|p{1.3cm}|}
\hline
$X_{23}$ & $\alpha_3$\\
$X_{31}$ & $\alpha_3$\\
$X_{32}$ & $\alpha_4$\\
$Z_{11}$ & $\alpha_2$\\
$Z_{22}$ & $\alpha_2$\\
$Z_{33}$ & $\alpha_2$\\
\hline
\end{tabular}
~
\begin{tabular}{|c|p{1.3cm}|}
\hline
$\Lambda_{12}$ & $\alpha_2 +\alpha_3$\\
$\Lambda_{13}$ & $\alpha_2 +\alpha_4$\\
$\Lambda_{21}$ & $\alpha_2 +\alpha_4$\\
$\Lambda_{23}$ & $\alpha_2 +\alpha_3$\\
$\Lambda_{31}$ & $\alpha_2 +\alpha_3$\\
$\Lambda_{32}$ & $\alpha_2 +\alpha_4$\\
\hline
\end{tabular}
~
\begin{tabular}{|c|p{1.3cm}|}
\hline
$\Lambda_{11}$ & $\alpha_1 +\alpha_2$\\
$\Lambda_{22}$ & $\alpha_1 +\alpha_2$\\
$\Lambda_{33}$ & $\alpha_1 +\alpha_2$\\
\hline
\end{tabular}
\caption{
Global symmetry charge assignments on chiral and Fermi fields of the $\mathbb{C}^{2}/\mathbb{Z}_{3} \times \mathbb{C}^2$ theory.
\label{t_global_3a}}
 \end{center}
 \end{table}

Let us consider the following mass deformation
\beal{es40a10}
\begin{array}{rrrclcrcl}
& & J & \textcolor{blue}{+}  & \textcolor{blue}{\Delta J} & & & E  \\
\Lambda_{11} : & \ \ \ & \textcolor{blue}{+ X_{11}} +  X_{13} \cdot X_{31}  &-&  X_{12} \cdot X_{21}  & \ \ \ \ & Z_{11} \cdot X_{11}  &-&  X_{11} \cdot  Z_{11} \\
\Lambda_{22} : & \ \ \ & \textcolor{blue}{-  X_{22}} + X_{21} \cdot X_{12}  &-&  X_{23} \cdot X_{32}  & \ \ \ \ & Z_{22} \cdot X_{22}  &-&  X_{22} \cdot  Z_{22} \\
\end{array}
 ~.~
\eea
Interestingly, in contrast with the previous examples, we only give mass to a subset of the chiral-Fermi pairs that can in principle become massive.

Integrating out the massive fields, $X_{11}$ and $X_{22}$ undergo a field replacement of the following form,
\beal{es40a11}
X_{11} & = - X_{13} \cdot X_{31} + X_{12} \cdot X_{21} ~,~ \nn\\
X_{22} & = + X_{21} \cdot X_{12} - X_{23} \cdot X_{32} ~.~
\eea
Furthermore, we note that the massive chiral fields $X_{11}$ and $X_{22}$ and the massive Fermi fields $\Lambda_{11}$ and $\Lambda_{22}$ are all part of the massive brick matching $p_1$, which is highlighted in blue in the $P$-matrix in \eref{es40a02}.

\begin{figure}[H]
\begin{center}
\resizebox{0.35\hsize}{!}{
\includegraphics[height=6cm]{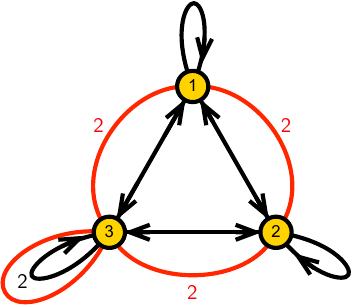} 
}
\caption{
The quiver obtained from the $\mathbb{C}^{2}/\mathbb{Z}_{3} \times \mathbb{C}^2$ theory after the mass deformation in \eqref{es40a10}.
\label{f_quiver_3b}}
 \end{center}
 \end{figure}

With the replacements in \eref{es40a11}, the $J$- and $E$-terms become,
\beal{es40a12}
\begin{array}{rcl}
 \Lambda_{12} : & J:  & \textcolor{BrickRed}{ X_{21} \cdot X_{12} \cdot X_{21}} - X_{23} \cdot X_{32} \cdot X_{21} -
 \textcolor{BrickRed}{ X_{21} \cdot X_{12} \cdot X_{21}} + X_{21} \cdot X_{13} \cdot X_{31}   \\
& E: & Z_{11} \cdot X_{12} - X_{12} \cdot Z_{22}   \\[.1cm]
 \Lambda_{13} : & J: & X_{31} \cdot X_{12} \cdot X_{21} - \textcolor{cyan}{X_{31} \cdot X_{13} \cdot X_{31} - X_{33} \cdot X_{31}}    \\
 & E: & Z_{42} \cdot X_{21} - X_{43} \cdot Z_{31}    \\[.1cm] 
 \Lambda_{21} : & J:  & \textcolor{BrickRed}{ X_{12} \cdot X_{21} \cdot X_{12}} - X_{12} \cdot X_{23} \cdot X_{32}  - \textcolor{BrickRed}{X_{12} \cdot X_{21} \cdot X_{12}} + X_{13} \cdot X_{31} \cdot X_{12}    \\
 & E: & Z_{22} \cdot X_{21} - X_{21} \cdot Z_{11}    \\[.1cm] 
 \Lambda_{23} : & J:  & \textcolor{cyan}{X_{33} \cdot X_{32}} - X_{32} \cdot X_{21} \cdot X_{12} + \textcolor{cyan}{X_{32} \cdot X_{23} \cdot X_{32}}    \\
 & E: & Z_{22} \cdot X_{23} - X_{23} \cdot Z_{33}   \\[.1cm] 
 \Lambda_{31} : & J:  & X_{12} \cdot X_{21} \cdot X_{13} - \textcolor{cyan}{X_{13} \cdot X_{31} \cdot X_{13}  - X_{13} \cdot X_{33}}    \\
  & E: & Z_{33} \cdot X_{31} - X_{31} \cdot Z_{11}    \\
 \Lambda_{32} : & J:  & \textcolor{cyan}{X_{23} \cdot X_{33}} - X_{21} \cdot X_{12} \cdot X_{23} + \textcolor{cyan}{X_{23} \cdot X_{32} \cdot X_{23}}   \\
 & E: & Z_{33} \cdot X_{32} - X_{32} \cdot Z_{22}   \\
\Lambda_{33} : & J:  & X_{32} \cdot X_{23} - X_{31} \cdot X_{13}   \\
  & E: & Z_{33} \cdot X_{33} - X_{33} \cdot Z_{33}   
\end{array}
~.~
\nn\\
\eea
Above, the terms in red cancel each other out trivially.
However, the terms in blue do not cancel out and break the toric condition of the $J$- and $E$-terms.
In order to restore the toric condition, we redefine the interaction terms as discussed in section \sref{sec:31}.
The following redefinition of the $J$-terms fixes the terms in \eref{es40a12},
\beal{es40a12b}
\Lambda_{13} \cdot X_{33} & \rightarrow  \Lambda_{13} \cdot (X'_{33} - X_{31} X_{13} ) ~,~\nn\\
\Lambda_{23} \cdot X_{33} & \rightarrow \Lambda_{23} \cdot (X'_{33} - X_{23} X_{32} ) ~,~\nn\\
X_{33} \cdot \Lambda_{31} & \rightarrow (X'_{33} - X_{31} X_{13} ) \cdot \Lambda_{31} ~,~\nn\\
X_{33} \cdot \Lambda_{32} & \rightarrow (X'_{33} - X_{32} X_{23} ) \cdot \Lambda_{32} ~.~
\eea
\fref{f_quiver_3b} shows the resulting quiver.

The $J$- and $E$-terms are,
\beal{es40a20}
\begin{array}{rrrclcrcl}
& & & J  & & & &  E   \\
\Lambda_{12 } : & \ \ \ &  X_{21} \cdot X_{13} \cdot X_{31}  &-&   X_{23} \cdot X_{32} \cdot X_{21}   & \ \ \ \ &Z_{11} \cdot X_{12}  &-&  X_{12} \cdot Z_{22}  \\
\Lambda_{21} : & \ \ \ & X_{13} \cdot X_{31} \cdot X_{12}  &-&  X_{12} \cdot X_{23} \cdot X_{32}   & \ \ \ \ & Z_{22} \cdot X_{21}  &-&  X_{21} \cdot Z_{11}  \\
\Lambda_{13} : & \ \ \ &  X_{31} \cdot  X_{12} \cdot X_{21}  &-&  X_{33} \cdot X_{31}   & \ \ \ \ & Z_{11} \cdot X_{13}  &-&  X_{13} \cdot  Z_{33}  \\
\Lambda_{31} : & \ \ \ & X_{12} \cdot X_{21} \cdot X_{13}   &-& X_{13} \cdot X_{33}   & \ \ \ \ & Z_{33} \cdot X_{31}  &-&  X_{31} \cdot Z_{11}   \\
\Lambda_{23} : & \ \ \ & X_{33} \cdot X_{32}  &-&  X_{32} \cdot X_{21} \cdot X_{12}   & \ \ \ \ & Z_{22} \cdot X_{23}  &-&  X_{23} \cdot Z_{33}  \\
\Lambda_{32} : & \ \ \ & X_{23} \cdot X_{33}  &-&  X_{21} \cdot X_{12} \cdot X_{23}   & \ \ \ \ &  Z_{33} \cdot X_{32}  &-&  X_{32} \cdot Z_{22}  \\
\Lambda_{33} : & \ \ \ & X_{32} \cdot X_{23}  &-&  X_{31} \cdot X_{13}   & \ \ \ \ & Z_{33} \cdot X_{33}  &-&  X_{33} \cdot  Z_{33}  
\end{array}
~.~
\eea
This theory has $(2,2)$ SUSY and corresponds to $\text{SPP} \times \IC$ \cite{Franco:2015tna}. It is straightforward to verify that the quiver diagram in \fref{f_quiver_3b} and the $J$- and $E$-terms in \eqref{es40a20} follow by dimensional reduction from the $4d$ $\mathcal{N}=1$ theory for SPP \cite{Morrison:1998cs}.

\begin{figure}[H]
\begin{center}
\resizebox{0.32\hsize}{!}{
\includegraphics[height=6cm]{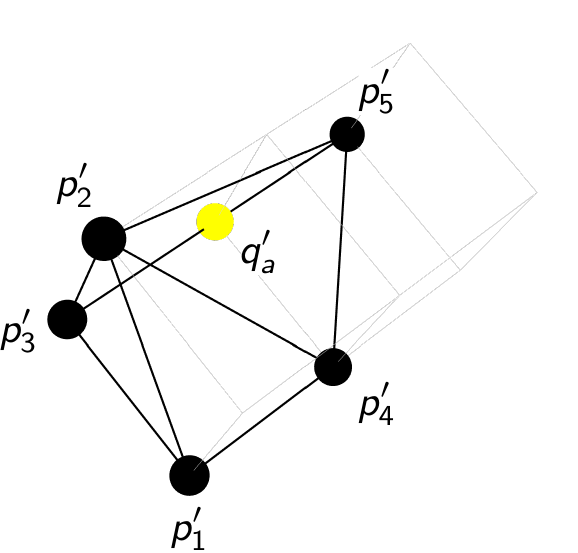} 
}
\caption{
The toric diagram for $\text{SPP} \times \IC$ obtained from a mass deformation of the $\mathbb{C}^{2}/\mathbb{Z}_{3} \times \mathbb{C}^2$ theory.
\label{f_toric_3b}}
 \end{center}
 \end{figure}

Using the forward algorithm, we can calculate the corresponding $P$-matrix, which takes the form
\beal{es40a21}
P_{\Lambda} = 
\resizebox{0.28\textwidth}{!}{$
\left(
\begin{array}{c|ccccc|cc}
 & p'_1 & p'_2 & p'_3 & p'_4 & p'_5 & q'_1 & q'_2  \\
\hline
X_{33} & 1 & 0 & 0 & 1 & 0 & 0 & 0 \\ 
X_{12} & 0 & 0 & 0 & 1 & 0 & 0 & 0 \\ 
X_{13} & 0 & 0 & 1 & 0 & 0 & 0 & 1 \\ 
X_{21} & 1 & 0 & 0 & 0 & 0 & 0 & 0 \\ 
X_{23} & 0 & 0 & 0 & 0 & 1 & 0 & 1 \\ 
X_{31} & 0 & 0 & 0 & 0 & 1 & 1 & 0 \\ 
X_{32} & 0 & 0 & 1 & 0 & 0 & 1 & 0 \\
Z_{11} & 0 & 1 & 0 & 0 & 0 & 0 & 0 \\
Z_{22} & 0 & 1 & 0 & 0 & 0 & 0 & 0 \\ 
Z_{33} & 0 & 1 & 0 & 0 & 0 & 0 & 0 \\
\hline
\Lambda_{12} & 0 & 1 & 0 & 1 & 0 & 0 & 0 \\
\Lambda_{13} & 0 & 1 & 1 & 0 & 0 & 0 & 1 \\
\Lambda_{21} & 1 & 1 & 0 & 0 & 0 & 0 & 0 \\
\Lambda_{23} & 0 & 1 & 0 & 0 & 1 & 0 & 1 \\
\Lambda_{31} & 0 & 1 & 0 & 0 & 1 & 1 & 0 \\ 
\Lambda_{32} & 0 & 1 & 1 & 0 & 0 & 1 & 0 \\
\Lambda_{33} & 1 & 1 & 0 & 1 & 0 & 0 & 0
\end{array}
\right) 
$}
~.~
\eea
We have labelled the new brick matchings $p'_\mu$ to distinguish them from the original ones. The new $Q_{JE}$ and $Q_{D}$ matrices are
\beal{es40a22}
Q_{JE} = 
\resizebox{0.25\textwidth}{!}{$
\left(
\begin{array}{ccccc|cc}
p'_1 & p'_2 & p'_3 & p'_4 & p'_5 & q'_1 & q'_2 \\
\hline
0 & 0 & -1 & 0 & -1 & 1 & 1
\end{array}
\right)
$}
~,~
Q_{D} = 
\resizebox{0.25\textwidth}{!}{$
\left(
\begin{array}{ccccc|cc}
p'_1 & p'_2 & p'_3 & p'_4 & p'_5 & q'_1 & q'_2 \\
\hline
-1 & 0 & 1 & 1 & 0 & -1 & 0 \\ 
1 & 0 & 0 & -1 & 1 & -1 & 0
\end{array}
\right) 
$} 
~.~
\eea
The corresponding toric diagram is summarized in
\beal{es40a23}
G_{t} = 
\resizebox{0.23\textwidth}{!}{$
\left(
\begin{array}{ccccc|cc}
p'_1 & p'_2 & p'_3 & p'_4 & p'_5 & q'_1 & q'_2 \\
\hline
1 & 1 & 1 & 1 & 1 & 1 & 1 \\ 
1 & 0 & 0 & 1 & 0 & 0 & 0 \\ 
0 & 0 & 0 & 1 & 2 & 1 & 1 \\ 
0 & 1 & 0 & 0 & 0 & 0 & 0
\end{array}
\right)
$}
~,~
\eea
and shown in \fref{f_toric_3b}. It indeed corresponds to $\text{SPP} \times \IC$.

\begin{figure}[H]
\begin{center}
\resizebox{0.8\hsize}{!}{
\includegraphics[height=6cm]{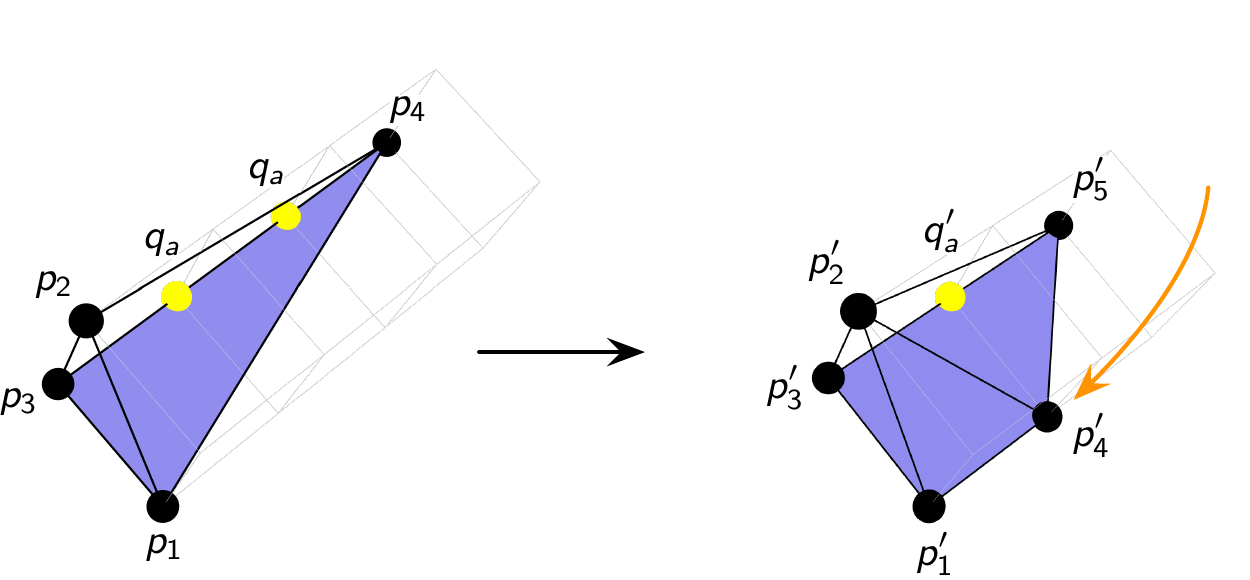} 
}
\caption{The toric diagram of $\mathbb{C}^{2}/\mathbb{Z}_{3} \times \mathbb{C}^2$  changing into the toric diagram of $\text{SPP} \times \IC$ under mass deformation. Brick matchings $p_1$, $p_3$ and $p_4$ define a 2-dimensional plane (blue) and correspond to the massive terms added to the $J$- and $E$-terms. On this plane, the point associated to $p_4$ moves to a new position due to the mass deformation.
\label{f_toric_3a3}}
 \end{center}
 \end{figure}

Comparing the toric diagrams before and after the mass deformation reveals that most of the points in them preserve their positions. Only the point corresponding to $p_4$ in $\mathbb{C}^{2}/\mathbb{Z}_{3} \times \mathbb{C}^2$ moves. In this process, some of the $q_a$'s become the single extremal brick matching $p'_5$, as  shown in \fref{f_toric_3a3}. The massive brick matching $p_1$, the unaffected brick matching $p_3$ and the moving brick matching $p_4$ define a plane as shown in \fref{f_toric_3a3}. When $p_4$ moves, it remains on this plane until it reaches its final location $p'_4$, as shown in \fref{f_toric_3a3}. 

Following the charge assignments in \tref{t_global_3a}, the mass terms in \eref{es40a10}, 
\beal{es40a25}
\Lambda_{11} \cdot X_{11} ~,~
\Lambda_{22} \cdot X_{22} ~,~
\eea
carry the following global symmetry charges,
\beal{es40a26}
\alpha(\Lambda_{11} \cdot X_{11} ) =  \alpha_1 + (\alpha_1 + \alpha_2) ~,~
\alpha(\Lambda_{22} \cdot X_{22} ) = \alpha_1 + (\alpha_1 + \alpha_2) ~,~
\eea
which by using the constraint $\alpha_1+\alpha_2+\alpha_3+\alpha_4 = 0 $ from \eref{es40a06b} can be re-written as
\beal{es40a26b}
\alpha(\Lambda_{11} \cdot X_{11} ) =  \alpha_1 - \alpha_3 - \alpha_4 ~,~
\alpha(\Lambda_{22} \cdot X_{22} ) = \alpha_1 - \alpha_3 - \alpha_4 ~.~
\eea
Once again all the mass deformation terms carry exactly the same non-zero global symmetry charge $\alpha_1 - \alpha_3 - \alpha_4$. 
Furthermore, these charges correspond exactly to the three extremal brick matchings $p_1$, $p_3$ and $p_4$ whose corresponding points in the toric diagram are on the mass deformation plane. The mass deformation break part of the global symmetry of the original theory $\mathbb{C}^{2}/\mathbb{Z}_{3} \times \mathbb{C}^2$.
Accordingly, we identify the mass deformation plane in \fref{f_toric_3a3} as the geometric directions along which the global symmetry of the 
theory $\mathbb{C}^{2}/\mathbb{Z}_{3} \times \mathbb{C}^2$ breaks. The global symmetry of $\text{SPP} \times \IC$ contains linear combinations of the original symmetries which are orthogonal to $\alpha_1 - \alpha_3 - \alpha_4$.

The minimum volumes of the underlying Sasaki-Einstein 7-manifolds are 
\beal{es40a30}
V_{min}^{\mathbb{C}^{2}/\mathbb{Z}_{3} \times \mathbb{C}^2}  = \frac{1}{3}  ~,~
V_{min}^{\text{SPP} \times \IC} = \frac{2}{3\sqrt{3}}  ~.~
\eea
Their ratio is
\beal{es40a31}
\frac{V_{min}^{\mathbb{C}^{2}/\mathbb{Z}_{3} \times \mathbb{C}^2}}{V_{min}^{\text{SPP} \times \IC}} 
= \frac{\sqrt{3}}{2} \simeq 0.866025 < 1 ~,~
\eea
which confirms that the minimum volume increases under mass deformation.

The mass deformation from $\mathbb{C}^{2}/\mathbb{Z}_{3} \times \mathbb{C}^2$ to $\text{SPP} \times \IC$ discussed above is related, via dimensional reduction, to the one connecting the $4d$ gauge theories for $\mathbb{C}^{2}/\mathbb{Z}_{3} \times \mathbb{C}$ and $\text{SPP}$. The connection between these two geometries can be observed on the mass deformation plane in \fref{f_toric_3a3}.
\\

\section{Conclusions and Future Directions \label{sec:5}}

In this paper we have focused on mass deformations connecting the $2d$ $(0,2)$ gauge theories associated to different toric CY 4-folds. These deformations generalize to $2d$ the Klebanov-Witten deformation that connects $4d$ gauge theories for the $\mathbb{C}^2/\mathbb{Z}_2 \times \mathbb{C}$ orbifold and the conifold.

\begin{figure}[ht!]
\begin{center}
\resizebox{0.6\hsize}{!}{
\includegraphics[height=6cm]{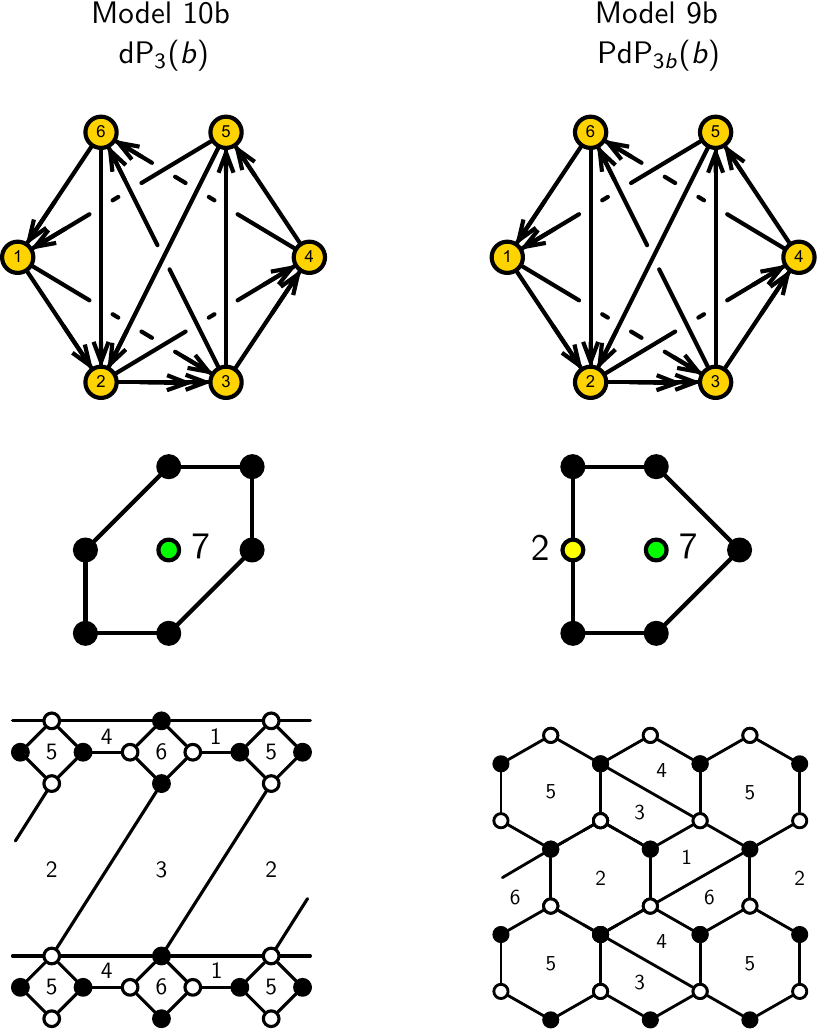} 
}
\caption{
Brane tilings for $\text{dP}_3$ phase $(b)$ (Model $10b$ in \cite{Hanany:2012hi}) and $\text{PdP}_{3b}$ phase $(b)$ (Model $9b$ in \cite{Hanany:2012hi}) \cite{Feng:2002zw, Franco:2005rj, Feng:2004uq, Hanany:2012hi}. Both theories have the same quiver and can be regarded as related by a non-mass relevant deformation.
\label{f_newdeform}}
 \end{center}
 \end{figure}

This class of deformations has interesting properties that follow from its connection to geometry. For instance, they translate into combinatorial transformations in terms of perfect matchings. Also, the pattern of symmetry breaking triggered by the deformation is captured by the modification of the underlying toric diagram.

We explored how the volume of the Sasaki-Einstein 7-manifold at the base of the CY 4-fold varies under deformation. Our results are consistent with it growing monotonically as energy is decreased, providing a natural distinction between UV and IR theories. The number of degrees of freedom of the gauge theory appears to be inversely proportional to the volume of the SE 7-manifold.

It is natural to ask whether there are more general relevant deformations, i.e. higher than quadratic order, that also have the special property of connecting toric CY 4-folds. Such deformations would preserve the quiver but modify the $J$- and $E$-terms. The analogous problem for $4d$ theories associated to toric CY 3-folds and brane tilings was studied in \cite{Cremonesi_to_appear}. An early example of this phenomenon, which was originally presented in \cite{Franco:2005rj}, corresponds to $dP_3$ and $PdP_3$. The toric diagrams, quivers and brane tilings for the two theories are shown in \fref{f_newdeform}. The superpotentials are,
\beal{es50a01}
W_{10b} &=& + X_{61} X_{13} X_{36} + X_{24} X_{46} X_{62} + X_{35} X_{52} X_{23}^{2} + X_{34} X_{45} X_{51} X_{12} X_{23}^{1}
\nn\\
&& - X_{24} X_{45} X_{52} - X_{36} X_{62} X_{23}^{1} - X_{35} X_{51} X_{13} - X_{12} X_{23}^{2} X_{34} X_{46} X_{61}
~,~
\\
W_{9b} &=& +X_{23}^{2} X_{36} X_{62} + X_{35} X_{52} X_{23}^{1} + X_{46} X_{61} X_{13} X_{34} + X_{24} X_{45} X_{51} X_{12} 
\nn\\
&& - X_{46} X_{62} X_{24} - X_{35} X_{51} X_{13} - X_{45} X_{52} X_{23}^{2} X_{34} - X_{23}^{1} X_{36} X_{61} X_{12}
~.~
\eea

We leave a detailed investigation of relevant deformations of toric CY$_4$’s for future work. The dimensional reductions of the $dP_3$ and $PdP_3$ pair, namely $dP_3 \times \mathbb{C}$ and $PdP_3\times \mathbb{C}$, provide an explicit example of this type of deformations. At present, there is no analogue of $a$-maximization for the $2d$ theories associated to CY$_4$’s, so a meticulous analysis similar to the one discussed above in $4d$ is not possible. While distinguishing between UV and IR theories is straightforward in theories related by turning on masses, it can become more challenging for non-mass deformations in the absence of such field theoretic tools. However, the connection to geometry can be used to identify the UV and IR theories since, as previously mentioned, it is expected that the volume of the underlying SE 7-manifold increases towards the IR.

\section*{Acknowledgements}

S.F. is supported by the U.S. National Science Foundation grants PHY-2112729 and DMS-1854179.
D.G. is supported by the Basic Science Research Program of the National Research Foundation of Korea (NRF) under the Ministry of Education in Korea (NRF-2022R1A6A3A03068148), and by JST PRESTO Grant Number JPMJPR2117.
D.G. would like to thank UNIST where this project was initiated.
G.G. is supported by a Simons Foundation grant for the Initiative for the Theoretical Sciences at the CUNY Graduate Center. 
R.-K. S. is supported by a Basic Research Grant of the National Research Foundation of Korea (NRF-2022R1F1A1073128).
He is also supported by a Start-up Research Grant for new faculty at UNIST (1.210139.01), a UNIST AI Incubator Grant (1.230038.01) and UNIST UBSI Grants (1.220123.01, 1.230065.01), as well as an Industry Research Project (2.220916.01) funded by Samsung SDS in Korea.  
He is also partly supported by the BK21 Program (``Next Generation Education Program for Mathematical Sciences'', 4299990414089) funded by the Ministry of Education in Korea and the National Research Foundation of Korea (NRF).
S.F. and R.-K. S. are grateful to the Simons Center for Geometry and Physics for hospitality during the project.


\bibliographystyle{JHEP}
\bibliography{mybib}

\end{document}